\documentclass[sigconf]{acmart}

\usepackage[normalem]{ulem}
 \usepackage{amsmath}
  \usepackage{multirow}
\usepackage{epstopdf}
  \usepackage{graphicx}
\usepackage{appendix}

\usepackage[linesnumbered,ruled,vlined]{algorithm2e}

\usepackage[normalem]{ulem}
\useunder{\uline}{\ul}{}
 \usepackage{enumitem}
 \useunder{\uline}{\ul}{}
 \newcommand{\para}[1]{\noindent \textbf{#1}}

\AtBeginDocument{%
  \providecommand\BibTeX{{%
    \normalfont B\kern-0.5em{\scshape i\kern-0.25em b}\kern-0.8em\TeX}}}

\newcommand{\chenc}[1]{\textcolor{black}{#1}}

\newtheorem{definition}{Definition}
\newtheorem{theorem}{Theorem}
\newtheorem{theorem1}{Theorem}

\copyrightyear{2024}
\acmYear{2024}
\setcopyright{rightsretained}
\acmConference[WSDM'24]{Proceedings of the 17th ACM International Conference on Web Search and Data Mining}{March 4th-8th, 2024}{Mérida, Yucatán, México}
\acmBooktitle{Proceedings of the 17th ACM International Conference on Web Search and Data Mining (WSDM'24), Mérida, Yucatán, México}
\acmDOI{10.1145/3543507.3583278}
\acmISBN{978-1-4503-9416-1/23/04}

\usepackage{etoolbox}
\makeatletter
\patchcmd{\maketitle}{\@copyrightpermission}{
   \begin{minipage}{0.3\columnwidth}
     \href{http://creativecommons.org/licenses/by/4.0/}{\includegraphics[width=0.90\textwidth]{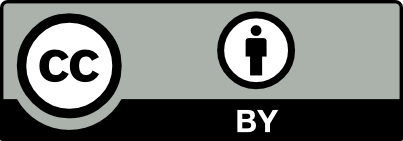}}
   \end{minipage}\hfill
   \begin{minipage}{0.7\columnwidth}
     \href{http://creativecommons.org/licenses/by/4.0/}{This work is licensed under a Creative Commons Attribution International 4.0 License.}
   \end{minipage}
  
   \vspace{5pt}
}{}{}

\author{Guanyu Lin$^{1}$, Chen Gao$^{2}$, Yu Zheng$^{2}$, Yinfeng Li$^{3}$, Jianxin Chang$^{3}$, Yanan Niu$^{3}$, Yang Song$^{3}$, Kun Gai$^{4}$, Zhiheng Li$^{2}$, Depeng Jin$^{2}$, Yong Li$^{2}$}
\affiliation{
\institution{$^1$Carnegie Mellon University, $^2$Tsinghua University, $^3$Kuaishou Technology, $^4$Unaffiliated}
 \institution{guanyul@andrew.cmu.edu, chgao96@gmail.com, zhengyu.davy@foxmail.com, \\ \{liyinfeng,  changjianxin, niuyanan, yangsong\}@kuaishou.com, \\ gai.kun@qq.com, \{zhhli, jindp, liyong07\}@tsinghua.edu.cn}
 \country{}
}

\renewcommand{\shortauthors}{Lin et al.}

\begin{document}

\title{Inverse Learning with Extremely Sparse Feedback for Recommendation}

\begin{abstract}
 
Modern personalized recommendation services often rely on user feedback, either explicit or implicit, to improve the quality of services. Explicit feedback refers to behaviors like ratings, while implicit feedback refers to behaviors like user clicks. However, in the scenario of full-screen video viewing experiences like Tiktok and Reels, the click action is absent, resulting in unclear feedback from users, hence introducing noises in modeling training.
Existing approaches on de-noising recommendation mainly focus on positive instances while ignoring the noise in a large amount of sampled negative feedback. In this paper, we propose a meta-learning method to annotate the unlabeled data from loss and gradient perspectives, which considers the noises in both positive and negative instances. Specifically, we first propose an \textit{Inverse Dual Loss} (IDL) to boost the true label learning and prevent the false label learning.
Then we further propose an \textit{Inverse Gradient} (IG) method to explore the correct updating gradient and adjust the updating based on meta-learning. Finally, we conduct extensive experiments on both benchmark and industrial datasets where our proposed method can significantly improve AUC by 9.25\% against state-of-the-art methods. Further analysis verifies the proposed inverse learning framework is model-agnostic and can improve a variety of recommendation backbones. The source code, along with the best hyper-parameter settings, is available at this link: \url{https://github.com/Guanyu-Lin/InverseLearning}.
\end{abstract}


\maketitle

\section{Introduction}\label{sec:intro}
\begin{figure}[t!]
	\begin{center}
\setlength\tabcolsep{1.5pt}

\begin{tabular}{cc}

 {\includegraphics[width=.48\linewidth]{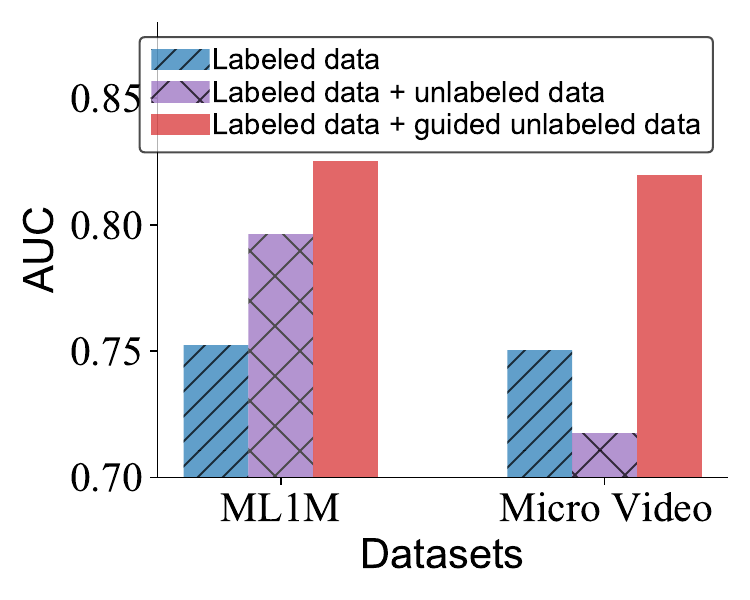}} &
 {\includegraphics[width=.48\linewidth]{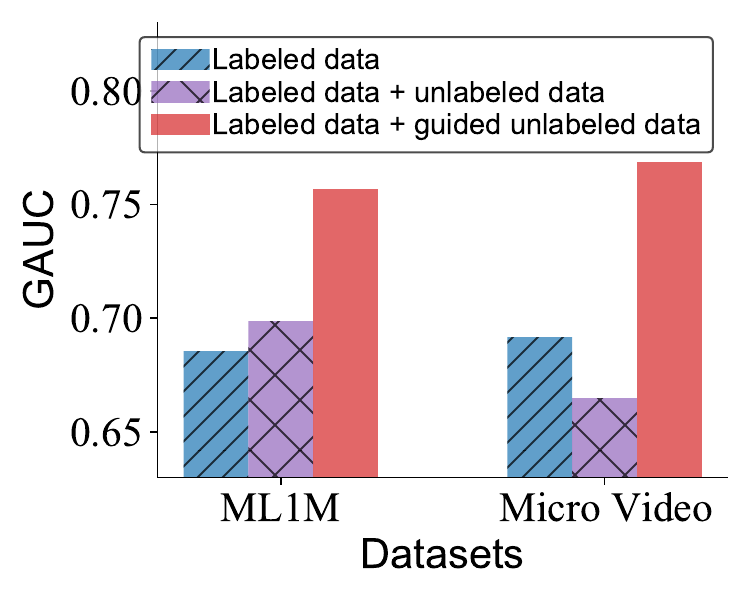}} 
\end{tabular}
	\end{center}
	\caption{Performance comparison of training with three data types under two datasets. Here "labeled data" means training with only labeled data; "labeled data and unlabeled data" means supervised training with labeled data and unsupervised training with unlabeled data; "labeled data mixed with guided unlabeled data" means splitting part of labeled data to guide the unsupervised learning on unlabeled data. }
	\label{fig:training}
\end{figure}

Recommender systems play a vital role in enhancing user experience and engagement, finding widespread use in online services like E-commerce and Micro-video platforms~\cite{ricci2011introduction, xue2017deep, liu2010personalized}. These systems aim to capture users' preferences based on their historical behaviors, with a focus on either explicit or implicit feedback.
Explicit feedback, such as user ratings, provides direct indications of user preferences but is challenging to collect due to the need for active user participation~\cite{jannach2018recommending, liang2021fedrec++}. In contrast, implicit feedback, including user clicks, purchases, and views, offers richer information and is more commonly utilized in modern recommender systems~\cite{liang2016modeling, chen2020efficient}.
In certain contexts like Micro-video platforms, users passively receive recommended items without actively engaging through actions like clicking or rating. Consequently, we encounter a scenario where the labeled feedback is extremely sparse, comprising predominantly quick-skip, long-stay, and a considerable number of slow-skip or short-stay videos with unclear feedback. Effectively leveraging this unlabeled feedback poses a significant challenge for recommendation systems.

The challenge of dealing with unclear feedback in recommender systems has led to various approaches that randomly sample unlabeled data and consider it as negative feedback, resulting in inevitable noise~\cite{he2017neural, chen2019social}. Typically, user-clicked data is treated as positive feedback, while unclicked data is sampled as negative feedback~\cite{he2017neural, chen2019social}. However, this sampling strategy may include positive instances in the unlabeled data, leading to false-negative cases. Additionally, some studies have explored hard negative sampling techniques, which reduce false-positive instances but increase false-negative instances~\cite{DNS, RNS, NEURIPS2020_0c7119e3}. Nevertheless, these methods often underperform when evaluated on true positive and negative data instead of the sampled negative data alone, as demonstrated in Table~\ref{tbl:hard_nega}. Notably, a recent work called DenoisingRec~\cite{Wang_2021} focuses on denoising positive feedback by manipulating the loss of false-positive instances but does not adequately address the issue of noisy negative feedback. Overall, existing approaches tend to concentrate solely on either the positive or negative perspective, without effectively tackling both aspects.

Our analysis of Figure~\ref{fig:training} reveals two key observations emerge, serving as the motivation behind our proposed method:

\begin{itemize}[leftmargin=*]
    \item \textbf{Full use of unlabeled data can boost the performance.} In the ML1M dataset, the introduction of unlabeled data boosts performance a lot. However, the performance is harmed in the Micro Video dataset, which means the solely unsupervised learning on unlabeled data is unstable.
    \item \textbf{Labeled data can guide the learning on unlabeled data.} In both two datasets, splitting part of labeled data to guide the unsupervised learning on unlabeled data can boost the performance. That is to say, guidance on unlabeled data can improve the robustness of unsupervised learning.
    
\end{itemize}

To simultaneously tackle the unclear feedback problem from positive and negative perspectives, we propose a novel learning-based approach that employs Inverse Dual Loss (IDL) and Inverse Gradient (IG). Our method automatically annotates the unlabeled data and subsequently adjusts the falsely annotated labels. As illustrated in Figure~\ref{fig:training}, we introduce IDL for unsupervised training on unlabeled data and leverage IG to guide the unlabeled data.

Specifically, the IDL is able to automatically annotate unlabeled data in an unsupervised learning fashion. The IDL employs a well-designed loss function that leverages both positive and negative feedback. We exploit the property that the loss associated with a false positive/negative instance exceeds that of a true positive/negative instance~\cite{Wang_2021}. By assigning different weights to the positive and negative labels of unlabeled instances, calculated using the inverse dual loss, we effectively utilize true positive/negative instances while mitigating the noise introduced by false positive/negative instances. This approach allows us to fully capitalize on valuable information and enhance the quality of annotation.

In addition, to adjust the false annotated labels and improve the robustness of IDL, we further propose an Inverse Gradient (IG) method. Here we build a meta-learning process~\cite{finn2017model, li2017meta} and split the training data into training-train and training-test data. We first exploit training-train data to pre-train the model. Then we further use training-test data to validate the correctness of classification by IDL. In other words, supervising the proposed unsupervised IDL method via training-test data.
Specifically, we calculate the gradient for the inverse dual loss of sampled instances as well as the additive inverse of the gradient. The model is optimized by either the direct gradient or the additive inverse of gradient, determined by the split training-test data. Experimental results illustrate that inverse gradient can truly improve the inverse dual loss.
In summary, the main contributions of this paper are as follows:

\begin{itemize}[leftmargin=*]
    \item We take the pioneering step to address the unclear passive feedback in video feed recommendation, which is far more challenging than existing works that are either based on explicit or implicit active feedback.
    \item We propose Inverse Dual Loss (IDL) to annotate the labels for sampled instances in an unsupervised learning manner. Besides, we further propose Inverse Gradient to guide the unsupervised learning on unlabeled data and improve the robustness of IDL.
    \item We experiment on two real-world datasets, verifying the superiority of our method compared with state-of-the-art approaches. Further studies sustain the effectiveness of our proposed method in label annotation and convergence. 
    
\end{itemize}

\section{Problem Definition}
We will formulate the problem here.
The recommendation task aims to model relevance score $\hat{y}^{\boldsymbol{\theta}}_{u i}=f(u, i | {\boldsymbol{\theta}})$ of user $u$ towards item $i$ under parameters ${\boldsymbol{\theta}}$. The \textit{LogLoss} function~\cite{DIN, DIEN} function to learn ideal parameters $\boldsymbol{\theta}^{*}$ is as:
\begin{equation}\label{eq:idea_loss}
\mathcal{L}_{\mathcal{D}^{*}}(\boldsymbol{\theta})  =
\frac{1}{\left|\mathcal{D}^{*}\right|}  \sum_{\left(u, i, y_{u i}^{*}\right) \in \mathcal{D}^{*}} 
\ell \left(\hat{y}^{\boldsymbol{\theta}}_{u i}, y_{u i}^{*}\right) ,
\end{equation}
where $\ell \left(\hat{y}^{\boldsymbol{\theta}}_{u i}, y_{u i}^{*}\right) = - \left( y_{u i}^{*} \log \left(\hat{y}^{\boldsymbol{\theta}}_{u i}\right)+\left(1-y_{u i}^{*}\right) \log \left(1-\hat{y}^{\boldsymbol{\theta}}_{u i}\right) \right)$, $y_{u i}^{*} \in\{0,1\}$ is the feedback of user $u$  towards item $i$. $\mathcal{D}^{*}$=$\left\{\left(u, i, y_{u i}^{*}\right)\right\},$ $u \in \mathcal{U}, i \in \mathcal{I}$ is the reliable interaction data between all user-item pairs.
Indeed, due to the limited collected feedback, the model training is truly formalized as follows: 
$\bar{\boldsymbol{\theta}}= \arg\min_{\boldsymbol{\theta}} \mathcal{L}_{\mathcal{D}^{l}}(\boldsymbol{\theta}) +  \mathcal{L}_{\mathcal{D}^{u}}(\boldsymbol{\theta})$,
where $\mathcal{D}^l\sim\mathcal{D}^{*}$ is the collected labeled data, and $\mathcal{D}^u =  \left\{\left(u, i, \bar{y}_{u i}\right) \mid u \in \mathcal{U}, i \in\mathcal{I}\right\}$ is the sampled unlabeled data where $\bar{y}_{u i} = 0$ is often assumed in existing recommenders for negative sampling. 
However, such a strategy will inevitably introduce noise because there are some positive unlabeled instances in the sampled data.
As a consequence, a model (i.e., $\bar{\boldsymbol{\theta}})$ trained with noisy data tends to exhibit suboptimal performance.
Thus, our goal is to construct a {denoising} recommender approximating to the ideal recommender ${\boldsymbol{\theta}}^{*}$ as:

\begin{equation}\label{eq:ideal_loss}
\boldsymbol{\theta}^{*}= \arg\min_{\boldsymbol{\theta}} \mathcal{L}_{\mathcal{D}^{l}}(\boldsymbol{\theta}) +  \mathcal{L}^{\text{denoise}}_{\mathcal{D}^{u}}(\boldsymbol{\theta}),
\end{equation}
where $\mathcal{L}^{\text{denoise}}_{\mathcal{D}^{u}}(\boldsymbol{\theta})$ indicates the loss on unlabeled data with all samples annotated correctly, \textit{i.e.} {denoising} sampling.

\section{Methodology}
In this section, we will first perform an in-depth analysis of existing solutions and their limitations. Then we will propose inverse dual loss to address the limitations of existing works for easy samples. Finally, we further propose inverse gradient to address the limitation of inverse dual loss and make it capable of not only easy samples but also hard samples that are misclassified.



\subsection{Inverse Dual Loss}
In this section, we first analyze the characteristics of existing solutions on the sampled unlabeled data. 
Then we introduce the proposed inverse dual loss solution to denoise sampled data.
\subsubsection{\textbf{Analysis of Existing Approach}}
\begin{figure*}[t]
	\begin{center}

\begin{tabular}{cccc}
\multicolumn{4}{c}{\includegraphics[height=0.7cm]{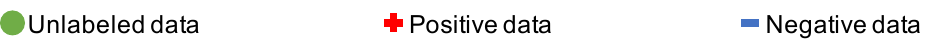}} \\

{\includegraphics[height=2.2cm]{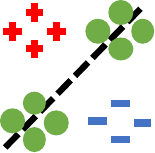}} & {\includegraphics[height=2.2cm]{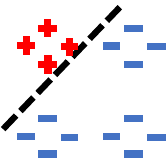}} & {\includegraphics[height=2.2cm]{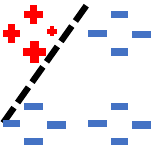}} & {\includegraphics[height=2.2cm]{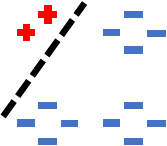}}\\
(a) Unlabeled data&
(b) Negative sampling&
(c) Reweight loss&
(d) Truncated loss \\
{\includegraphics[height=2.2cm]{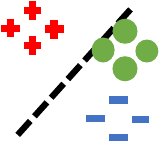}} & {\includegraphics[height=2.2cm]{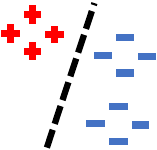}} & {\includegraphics[height=2.2cm]{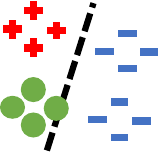}} & {\includegraphics[height=2.2cm]{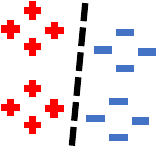}}\\
(e) \chenc{Easy negatives}&
(f) True-negative label&
(g) \chenc{Easy positives} &
(h) True-positive label\\
{\includegraphics[height=2.2cm]{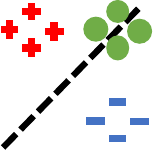}} & {\includegraphics[height=2.2cm]{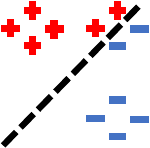}} & {\includegraphics[height=2.2cm]{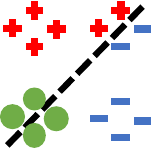}} & {\includegraphics[height=2.2cm]{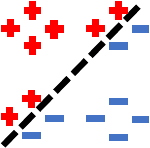}}\\
(i) \chenc{Hard negatives} &
(j) False-positive label&
(k) \chenc{Hard positives}&
(l)  False-negative label
\end{tabular}
	\end{center}
	\caption{Illustrations of existing solutions and our inverse dual loss's effectiveness and limitation.\textbf{ (a)-(d) are the illustrations of existing solutions}: (a) illustrates there are a lot of unlabeled data; (b) illustrates the traditional negative sampling approach; (c) illustrates the reweighted loss of DenoisingRec; (d) illustrates the truncated loss of DenoisingRec adapted on the false-negative instance. \textbf{(e)-(f) are the illustrations of our inverse dual loss's effectiveness with \chenc{easy sampling}}: (e) illustrates the easy negative instances are sampled; (f) illustrates labeling the sampled instances as true negative; (g) illustrates the easy positive instances are sampled; (h) illustrates labeling the sampled instances as true positive and approximate to \textbf{ground-truth}. \textbf{(i)-(l) are the illustrations of our inverse dual loss's limitation with \chenc{hard sampling}}: (i) illustrates the hard negative instances are sampled; (j) illustrates labeling part of the sampled instances as false positive; (k)  illustrates the hard positive instances are sampled; (l) illustrates labeling part of the sampled instances as false negative. }
	\label{fig:intuition}
\end{figure*}
We first explain the data sparsity problem in recommender systems from the perspective of classification boundary, based on which we will introduce the existing solutions. 
As shown in Figure~\ref{fig:intuition} (a), in recommender systems, labeled data tends to be extremely sparse compared with a large number of unlabeled data.
A recommendation model is prone to overfitting if it is only trained based on the sparsely labeled data, compared with the ground-truth in Figure~\ref{fig:intuition} (h).

In practice, existing recommenders often sample from unlabeled data and treat all the sampled data as negative feedback.
Such an approach introduces false negative, which fails to retrieve items that users may be interested in, as shown by the classification boundary in Figure~\ref{fig:intuition} (b).
That is, there exists noise in the sampled negative data.
However, existing denoising approaches mainly focus on the noise in positive samples (false positive).
For example, DenoisingRec \cite{Wang_2021} attempts to achieve denoising for false positive instances as:
\begin{equation}
\boldsymbol{\theta}^{*}= \arg\min_{\boldsymbol{\theta}} \mathcal{L}^{\text{denoise}}_{\mathcal{D}^{l} \cup \mathcal{D}^{noise}}(\boldsymbol{\theta}) +  \mathcal{L}_{\mathcal{D}^{u}}(\boldsymbol{\theta})
\end{equation}
where $\mathcal{D}^{noise} =  \left\{\left(u, i, 1\right) \mid u \in \mathcal{U}, i \in\mathcal{I}, y^{*}_{ui} = 0 \right\}$ is the noisy false positive data they introduce in experiments. 
For example, R-CE (Reweight Cross-Entropy) of DenoisingRec assigns lower weight on false-positive instances with large loss (Figure~\ref{fig:intuition} (c)), and T-CE (Truncated Cross-Entropy) of DenoisingRec discards those false positive instances with large loss (Figure~\ref{fig:intuition} (d)). 
Though achieving denoising for false-positive instances, they ignore false-negative instances and fail to address the noise brought by the negative sampling.

In fact, the collected labeled data is much cleaner than sampled unlabeled data, and the number of false-positive instances is limited in real-world recommender systems.
On the contrary, the noise brought by negative sampling is far more harmful.
In other words, the noise level of positive unlabeled data incorrectly sampled as negative feedback is much higher than that of negative samples wrongly regarded as positive feedback.

To sum up, existing solutions either introduce noise or perform incomplete denoising, which motivates us to further propose a {denoising} solution for unlabeled data from both positive and negative perspectives. 


\subsubsection{\textbf{Labeling with Inverse Dual Loss}}
As an existing attempt in DenoisingRec, we have discovered that the false positive instances are with a greater loss. It is also an apparent phenomenon in machine learning. For example, {if we have a positive instance and a well-trained model, the loss of classifying it as negative will be greater than that of classifying it as positive}. Otherwise, if we have a negative instance and a well-trained model, the loss of classifying it as positive will be greater. Hence, we can assume the sampled unlabeled instances are both possibly positive and negative and then exploit this inherent characteristic to automatically weigh more on the true positive or negative instances while weighing less on the false ones. 

\begin{figure*}[t!]
	\begin{center}
\begin{tabular}{cc}
\multicolumn{2}{c}{\includegraphics[height=0.6cm]{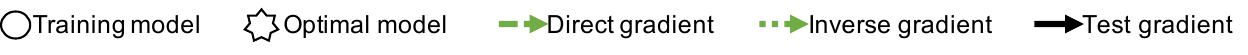}} \\ 
 {\includegraphics[height=1.5cm]{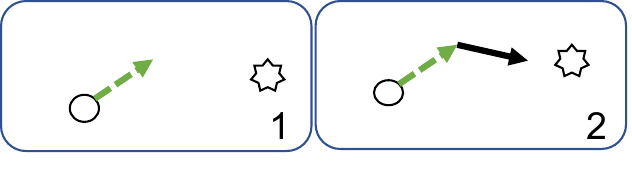}} & {\includegraphics[height=1.5cm]{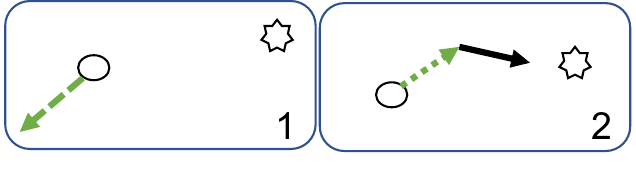}}\\
(a) Update with direct gradient &
(b) Update with inverse gradient \\
 {\includegraphics[height=1.3cm]{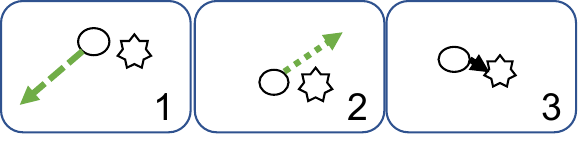}} &
 {\includegraphics[height=1.3cm]{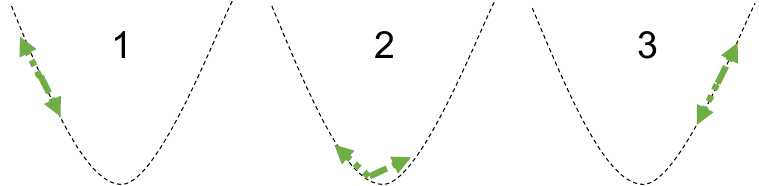}}\\
(c) Pass this batch of unlabeled data & (d) Convergence analysis
\\
\end{tabular}
	\end{center}
	\caption{Illustrations of inverse gradient adaption. }
	\label{fig:inverse_grad}
\end{figure*}
\begin{definition} (Inverse Dual Loss)
The inverse dual loss is defined as denoising loss to automatically classify the unlabeled data as:
\begin{equation}\label{eq:dual_loss}
\mathcal{L}^{dual}_{\mathcal{D}^u}(\boldsymbol{\theta}) = \frac{1}{\left|\mathcal{D}^u\right|} \sum_{\left(u, i, \bar{y}_{u i}\right) \in \mathcal{D}^u} w^1 \ell\left(\hat{y}^{\boldsymbol{\theta}}_{u i}, 1\right) + w^0 \ell\left(\hat{y}^{\boldsymbol{\theta}}_{u i}, 0\right)    
\end{equation}
where $w^1 = \frac{\textbf{stopgrad}\left(\ell\left(\hat{y}^{\boldsymbol{\theta}}_{u i}, 0\right)\right)}{z_w\textbf{stopgrad}\left(\ell\left(\hat{y}^{\boldsymbol{\theta}}_{u i}, 1\right)\right)}, w^0 = \frac{\textbf{stopgrad}\left(\ell\left(\hat{y}^{\boldsymbol{\theta}}_{u i}, 1\right)\right)}{z_w\textbf{stopgrad}\left(\ell\left(\hat{y}^{\boldsymbol{\theta}}_{u i}, 0\right)\right)}$ are the weights for positive loss and negative loss, respectively. $z_w = \frac{\textbf{stopgrad}\left(\ell\left(\hat{y}^{\boldsymbol{\theta}}_{u i}, 0\right)\right)}{\textbf{stopgrad}\left(\ell\left(\hat{y}^{\boldsymbol{\theta}}_{u i}, 1\right)\right)} + \frac{\textbf{stopgrad}\left(\ell\left(\hat{y}^{\boldsymbol{\theta}}_{u i}, 1\right)\right)}{\textbf{stopgrad}\left(\ell\left(\hat{y}^{\boldsymbol{\theta}}_{u i}, 0\right)\right)}$ is the normalization parameter. $\mathcal{D}^u$ is the sampled unlabeled data and $\boldsymbol{\theta}$ is the model parameters to be learned. Here \textbf{stopgrad} is a stop-gradient operation.
\end{definition}



The pros of inverse dual loss are as shown in (e)-(h) of Figure~\ref{fig:intuition}: (e) when the easy negative instances are sampled, the loss of classifying them as positive will be greater than that of classifying them as negative, and thus inverse dual loss will assign more weights on the negative loss; (f) gradually assigning more and more weights on the negative loss, the negative unlabeled instances will eventually be classified as negative; (g) likewise, when the easy positive instances are sampled, the inverse dual loss will assign more weights on the positive loss; (h) the positive unlabeled instances will eventually be classified as positive and approximate to the ground-truth.

\subsubsection{\textbf{Limitation of Inverse Dual Loss}}
When sampling the easy positive or negative instances, our inverse dual loss can boost learning by correctly labeling the sampled instances. However, it may be an obstacle when there are some \chenc{hard positive or negative instances}.
As shown in (i)-(l) of Figure~\ref{fig:intuition}, given some hard positive or negative instances, the classification boundary will be prevented from the ground-truth: (i) the hard negative instances are sampled; (j) half of the negative unlabeled instances will be classified as positive and become noise; (k) the hard positive instances are sampled; (l) half of the positive unlabeled instances will be classified as negative and become noise. That is to say, our inverse dual loss relies heavily on the current training classification boundary and the difficulty of sampled data, requiring us to further improve its robustness.

\subsection{Inverse Gradient}
To improve the robustness of Inverse Dual Loss on the hard sampled data towards the current training model, in this section, we further propose inverse gradient to adjust the gradient of false annotated data, inspired by the meta-learning framework~\cite{finn2017model, li2017meta}. Then we analyze the convergence of the proposed Inverse Gradient.
\subsubsection{\textbf{Learning to Label with Inverse Gradient}}
In this part, we introduce our solution for tackling the false annotated instances of Inverse Dual Loss. 
\begin{definition} \label{def:direct_grad}
(Inverse Gradient) We define the gradient and additive inverse of gradient calculated by \eqref{eq:dual_loss} w.r.t.$\nabla \mathcal{L}^{dual}_{\mathcal{D}^u}(\boldsymbol{\theta})$ and $- \nabla \mathcal{L}^{dual}_{\mathcal{D}^u}(\boldsymbol{\theta})$ as direct gradient and inverse gradient, respectively, of the loss for unlabeled data $\mathcal{D}^u$.
\end{definition}

\begin{theorem} \label{def:inverse_grad}
Given learning rate $\alpha \in \mathbb{R}, \alpha \neq 0$, assume the temporal model parameters updated by the direct gradient and inverse gradient, respectively, are as $\boldsymbol{\theta}^d = \boldsymbol{\theta} - \alpha \circ \nabla \mathcal{L}^{dual}_{\mathcal{D}^u}(\boldsymbol{\theta})$ and $\boldsymbol{\theta}^i = \boldsymbol{\theta} + \alpha \circ \nabla \mathcal{L}^{dual}_{\mathcal{D}^u}(\boldsymbol{\theta})$. Then, the relationship between the loss of them and the model with parameter $\boldsymbol{\theta}$ on data $\mathcal{D}^l$ will be either $\mathcal{L}_{\mathcal{D}^l}(\boldsymbol{\theta}^d) > \mathcal{L}_{\mathcal{D}^l}(\boldsymbol{\theta}) >  \mathcal{L}_{\mathcal{D}^l}(\boldsymbol{\theta}^i) $ or $ \mathcal{L}_{\mathcal{D}^l}(\boldsymbol{\theta}^d) < \mathcal{L}_{\mathcal{D}^l}(\boldsymbol{\theta}) < \mathcal{L}_{\mathcal{D}^l}(\boldsymbol{\theta}^i)$.
\end{theorem}

The proof of Theorem~\ref{def:inverse_grad} is as Appendix~\ref{appendix::proof}. Based on this theorem, we can have the following gradient updating strategies.
Generally, we will first split the training data into training-train data and training-test data, where we pre-train the model on the training-train data. Then we will calculate the direct gradient of inverse dual loss on the sampled unlabeled data, which can further result in the following three cases: 
\begin{itemize}[leftmargin=*]
    \item When the sampled data is easy, we can exploit the direct gradient to update the model, and it will gain a smaller test loss on the training-test data, as shown in Figure~\ref{fig:inverse_grad} (a);
    \item When the sampled data is hard, exploiting the inverse gradient to update the model will gain a smaller test loss on the training-test data, and thus we exploit inverse gradient here as shown in Figure~\ref{fig:inverse_grad} (b); 
    \item When the model is approximately optimal, either direct gradient or inverse gradient will prevent it from ground-truth, and we discard this batch of unlabeled data as shown in Figure~\ref{fig:inverse_grad} (c);
\end{itemize}


\begin{algorithm} 
\small
\SetKwData{Left}{left}\SetKwData{This}{this}\SetKwData{Up}{up} \SetKwFunction{Union}{Union}\SetKwFunction{FindCompress}{FindCompress} \SetKwInOut{Input}{input}\SetKwInOut{Output}{output}
	
	\Input{Labeled data $\mathcal{D}^l$, unlabeled data $\mathcal{D}^u$, learning rate $\gamma$, $\alpha$} 
	\Output{$\boldsymbol{\theta}$}
	 \BlankLine 
	 
	 \emph{Initialize $\boldsymbol{\theta}$, split $\mathcal{D}^l$ into training-train data $\operatorname{train}\left(\mathcal{D}^l\right)$ and training-test data $\operatorname{test}\left(\mathcal{D}^l\right)$}\; 
	 \While{not done}{
	 \For{$t= 1$ \KwTo $T$}{ 
	    
	 	\emph{Sample batch of training labeled data $\textit{train}(\mathcal{D}^l_t)\sim\mathcal{D}^l$}\; 

	 	\emph{$\mathcal{L}_{\operatorname{train}\left(\mathcal{D}^l_{t}\right)}(\boldsymbol{\theta}) = \frac{1}{\left|\operatorname{train}\left(\mathcal{D}^l_{t}\right)\right|} \sum_{(u, i, y^{*}_{ui}) \in \operatorname{train}\left(\mathcal{D}^l_{t}\right)} \ell\left(\hat{y}^{\boldsymbol{\theta}}_{ui}, \hat{y}^{*}_{ui}\right)$}\; 
	 	
	 	\emph{$\boldsymbol{\theta} = \boldsymbol{\theta}-\gamma \circ \nabla \mathcal{L}_{\operatorname{train}\left(\mathcal{D}_{t}\right)}(\boldsymbol{\theta})$}\;
	 	}
	 \For{$t= 1$ \KwTo $T$}{ 

	 \emph{Sample batch of training unlabeled data $\textit{train}(\mathcal{D}^u_t)\sim\mathcal{D}^u$}\; 
	 
	 \emph{$\mathcal{L}^{dual}_{\operatorname{train}\left(\mathcal{D}^u_{t}\right)}(\boldsymbol{\theta}) = \frac{1}{\left|\operatorname{train}\left(\mathcal{D}^u_{t}\right)\right|} \sum_{\left(u, i, \bar{y}_{u i}\right) \in \operatorname{train} \left(\mathcal{D}^u_{t}\right)} w^1 \ell\left(\hat{y}^{\boldsymbol{\theta}}_{u i}, 1\right) + w^0 \ell\left(\hat{y}^{\boldsymbol{\theta}}_{u i}, 0\right)$}\; 
	 	
 	\emph{$\boldsymbol{\theta}^{d} = \boldsymbol{\theta}-\alpha \circ \nabla \mathcal{L}^{dual}_{\operatorname{train}\left(\mathcal{D}^u_{t}\right)}(\boldsymbol{\theta})$}\;
 	
	 \emph{$\boldsymbol{\theta}^{i} = \boldsymbol{\theta} + \alpha \circ \nabla \mathcal{L}^{dual}_{\operatorname{train}\left(\mathcal{D}^u_{t}\right)}(\boldsymbol{\theta})$}\;
	 
	 \emph{Sample batch of test labeled data $\textit{test}(\mathcal{D}^l_t) \sim \mathcal{D}^l$}\; 

	 \emph{$\mathcal{L}_{\operatorname{test}\left(\mathcal{D}^l_{t}\right)}(\boldsymbol{\theta}^{d}) = \frac{1}{\left|\operatorname{test}\left(\mathcal{D}^l_{t}\right)\right|} \sum_{\left(u, i, y_{u i}^{*}\right) \in \operatorname{test}\left(\mathcal{D}^l_{t}\right)} \ell\left(\hat{y}^{\boldsymbol{\theta}^{d}}_{u i}, y_{u i}^{*}\right)$}\; 
	 
	 \emph{$\mathcal{L}_{\operatorname{test}\left(\mathcal{D}^l_{t}\right)}(\boldsymbol{\theta}^{i}) = \frac{1}{\left|\operatorname{test}\left(\mathcal{D}^l_{t}\right)\right|} \sum_{\left(u, i, y_{u i}^{*}\right) \in \operatorname{test}\left(\mathcal{D}^l_{t}\right)} \ell\left(\hat{y}^{\boldsymbol{\theta}^{i}}_{u i}, y_{u i}^{*}\right)$}\; 
	 
	 \emph{$\mathcal{L}_{\operatorname{test}\left(\mathcal{D}^l_{t}\right)}(\boldsymbol{\theta}) = \frac{1}{\left|\operatorname{test}\left(\mathcal{D}^l_{t}\right)\right|} \sum_{\left(u, i, y_{u i}^{*}\right) \in \operatorname{test}\left(\mathcal{D}^l_{t}\right)} \ell\left(\hat{y}^{\boldsymbol{\theta}}_{u i}, y_{u i}^{*}\right)$}\; 
	 
	 \emph{$\boldsymbol{\theta} = \arg\min_{\{\boldsymbol{\theta}^{d}, \boldsymbol{\theta} , \boldsymbol{\theta}^{i}\}}\{\mathcal{L}_{\operatorname{test}\left(\mathcal{D}^l_{t}\right)}(\boldsymbol{\theta}^{d}),  \mathcal{L}_{\operatorname{test}\left(\mathcal{D}^l_{t}\right)}(\boldsymbol{\theta}), \mathcal{L}_{\operatorname{test}\left(\mathcal{D}^l_{t}\right)}(\boldsymbol{\theta}^{i})\}$}\;
	 
	 \emph{$\mathcal{L}_{\operatorname{test}\left(\mathcal{D}^l_{t}\right)}(\boldsymbol{\theta}) = \min \{\mathcal{L}_{\operatorname{test}\left(\mathcal{D}^l_{t}\right)}(\boldsymbol{\theta}^{d}),  \mathcal{L}_{\operatorname{test}\left(\mathcal{D}^l_{t}\right)}(\boldsymbol{\theta} ), \mathcal{L}_{\operatorname{test}\left(\mathcal{D}^l_{t}\right)}(\boldsymbol{\theta}^{i})\}$}\;
	 
	 \emph{$\boldsymbol{\theta} = \boldsymbol{\theta}-\gamma \circ \nabla \mathcal{L}_{\operatorname{test}\left(\mathcal{D}^l_{t}\right)}(\boldsymbol{\theta})$}\;
	 
 	 } }
 	 	  \caption{Inverse Gradient Adaptation}
 	 	  \label{alg:ige} 
 	 \end{algorithm}

\subsubsection{\textbf{Algorithm}}\label{sec:alg}
We present the procedure of exploring these three cases as Algorithm~\ref{alg:ige}. The algorithm first pre-trains the model using the split training-train data. Then the model will update with direct gradient or inverse gradient or even not update, determined by the validation on the split training-test data.

More specifically, the inputs of our proposed algorithm are labeled data $\mathcal{D}^l$, unlabeled data $\mathcal{D}^u$, and learning rate $\gamma$, $\alpha$. The first iteration aims to pre-train the model using the split training-train data. The second iteration aims to explore the direct gradient and inverse gradient on the loss for sampled unlabeled data, where three strategies are explored here as lines 13-15 of Algorithm~\ref{alg:ige} with $\boldsymbol{\theta}^d, \boldsymbol{\theta}^i$ and $\boldsymbol{\theta}$ as the model parameters updated by the direct gradient, inverse gradient and without being updated by the gradient on the loss for sampled unlabeled data, respectively. Finally, the explored updated direction with minimal test loss on the training-test data will be selected to update the model for this iteration.

\subsubsection{\textbf{Convergence Analysis}}\label{sec:convergence}
As shown in Figure~\ref{fig:inverse_grad} (d), the first case illustrates when the unlabeled data is ideally sampled, the updating with direct gradient will lead to a smaller loss and better convergence on the training-test data, while the third case with poorly sampled data supposes to update with inverse gradient. However, when the model approximates convergence on the training-test data, updating with either direct gradient or inverse gradient may be poorer than no updating, as shown in stage 2 of Figure~\ref{fig:inverse_grad} (d). 

To avoid the gradient ascent problem for the second case, we can set the learning rate for the inverse dual loss to be smaller than that for the test loss, i.e., $\alpha < \gamma$ in Algorithm~\ref{alg:ige}. In this way, the scale of updating by the gradient for inverse dual loss will be within the scale of updating by the gradient for test loss. That is to say, the gradient ascent problem is less likely to occur on the inverse dual loss for unlabeled data than the loss for labeled data. 
\section{Experiments}
In this section, we perform experiments on two real-world datasets, targeting four research questions (RQs): 
\begin{itemize}[leftmargin=*]
	\item \textbf{RQ1:} How does the proposed method perform compared with the state-of-the-art denoising recommenders? What is the effect of two proposed components, i.e., Inverse Gradient (IG) and Inverse Dual Loss (IDL)?
		\item \textbf{RQ2:}  How does our proposed inverse dual loss identify the unlabeled data?
\item \textbf{RQ3:}  What is the effect of the inverse gradient on convergence?
  \item \textbf{RQ4:}  What is the optimal ratio between the learning rates for inverse dual loss and training-test?


\end{itemize}

We also study RQ5: "How does the proposed method perform compared with the state-of-the-art hard negative sampling recommenders?" in Appendix~\ref{appendix::hardNegCom}.
\subsection{Experimental Setup}
\subsubsection{\textbf{Datasets}}
 To practice and verify the effectiveness of our proposed method, we conduct experiments on an industrial Micro Video dataset and a public benchmark ML1M dataset, which is widely used in existing work for recommender systems~\cite{lian2018xdeepfm, cheng2020adaptive}. Micro Video is an extremely sparse dataset where users are passive in receiving the feed videos and have rare active feedback. We introduce the details of them in Appendix~\ref{appendix::dataset}.

\subsubsection{\textbf{Baselines and Evaluation Metrics}} To demonstrate the effectiveness of our proposed inverse learning on unlabeled data, we compared the performance of recommenders trained by our inverse gradient (IG) with recommenders trained by inverse dual loss (IDL) and normal training by standard loss or negative sampling (NS)~\cite{rendle2012bpr, he2017neural, chen2019social}. Besides we also compare our inverse learning method with the state-of-the-art methods for denoising recommender systems.
Specifically, we also compare two adaptive denoising training strategies, T-CE and R-CE, of DenoisingRec~\cite{Wang_2021}. Following DenoisingRec~\cite{Wang_2021}, we select GMF and NeuMF~\cite{he2017neural} as backbones, which are neural Collaborative Filtering models. The details of them are illustrated in Appendix~\ref{appendix::baselines}.


We adopt widely-used AUC and GAUC as accuracy metrics~\cite{gunawardana_evaluating_2015}. Besides, two widely-used ranking metrics~\cite{SURGE}, MRR and NDCG@10, are also adopted for evaluation. 

\begin{table}[t!]

\caption{Performance comparisons with GMF and NeuMF backbones on two datasets. Bold and underline refer to the best and second best results, respectively. {Here IG includes the IDL method}.}\label{tbl:overall}
\centering
\setlength\tabcolsep{1.5pt}
\begin{tabular}{cccccccc}
\hline
   \multirow{2}{*}{\textbf{Model}} & \multirow{2}{*}{\textbf{Method}} & \multicolumn{3}{c}{\textbf{ML1M}}                   & \multicolumn{3}{c}{\textbf{Micro Video}}            \\ \cline{3-8} 
                                &                                  & \textbf{NDCG}   & \textbf{AUC}    & \textbf{GAUC}   & \textbf{NDCG}   & \textbf{AUC}    & \textbf{GAUC}   \\ \hline
\multirow{6}{*}{\textbf{GMF}}   & \textbf{None}                    & 0.9285          & \underline{0.7671}         & 0.6919          & \underline{0.7365}          & \underline{0.8024}          & \underline{0.7558}          \\ \cline{2-8} 
                                & \textbf{NS}                      & \underline{0.9400}          & 0.7639          & 0.7200          & 0.7049          & 0.7802          & 0.7247          \\ \cline{2-8} 
                                & \textbf{T-CE}                    & {0.9349}          & 0.7612          & 0.7163          & 0.6994          & 0.6486          & 0.6956          \\ \cline{2-8} 
                                & \textbf{R-CE}                    & 0.9391          & 0.7632          & \underline{0.7192}          & 0.7069          & 0.7820          & 0.7260          \\ \cline{2-8} 
                                & \textbf{IDL}                     & 0.8996          & 0.7304          & 0.6400          & 0.7272          & 0.7858          & 0.7335          \\ \cline{2-8} 
                                & \textbf{{IG}}                      & \textbf{0.9521} & \textbf{0.8318} & \textbf{0.7642} & \textbf{0.7773} & \textbf{0.8033} & \textbf{0.7593} \\ \hline
\multirow{6}{*}{\textbf{NeuMF}} & \textbf{None}                    & 0.9214          & 0.7524          & 0.6856          & 0.6649          & \underline{0.7504}          & \underline{0.6916}          \\ \cline{2-8} 
                                & \textbf{NS}                      & 0.9298          & 0.7495          & 0.7088          & 0.6350          & 0.7191          & 0.6689          \\ \cline{2-8} 
                                & \textbf{T-CE}                    & 0.9351          & 0.7587          & \underline{0.7158}          & \underline{0.6725}          & 0.7469          & 0.6911          \\ \cline{2-8} 
                                & \textbf{R-CE}                    & \underline{0.9349}          & 0.7521          & 0.7117          & 0.6288          & 0.7038          & 0.6568          \\ \cline{2-8} 
                                & \textbf{IDL}                     & 0.9212          & \underline{0.7962}          & 0.6984          & 0.6466          & 0.7174          & 0.6647          \\ \cline{2-8} 
                                & \textbf{{IG}}                      & \textbf{0.9449} & \textbf{0.8253} & \textbf{0.7569} & \textbf{0.7809} & \textbf{0.8198} & \textbf{0.7689} \\ \hline
\end{tabular}
\end{table}

\begin{figure*}[t!]
	\begin{center}
\begin{tabular}{cccc}

 {\includegraphics[height=3.5cm]{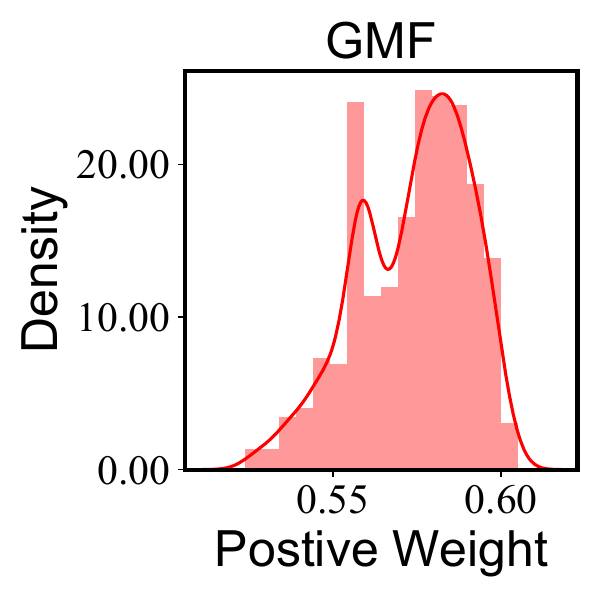}} &
 {\includegraphics[height=3.5cm]{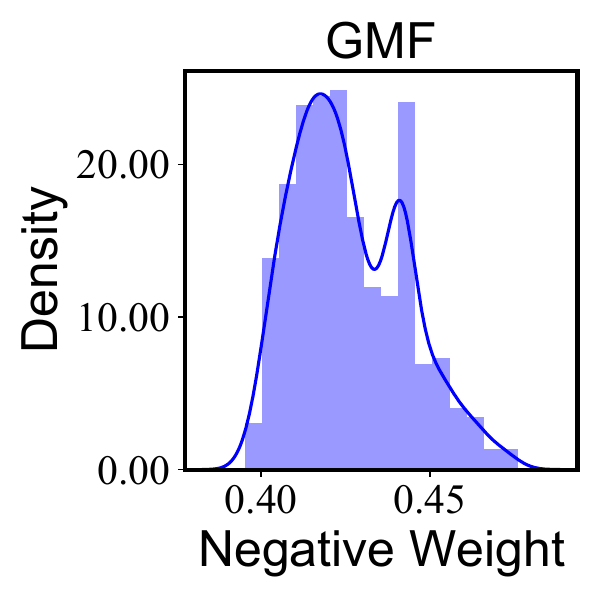}} &
 {\includegraphics[height=3.5cm]{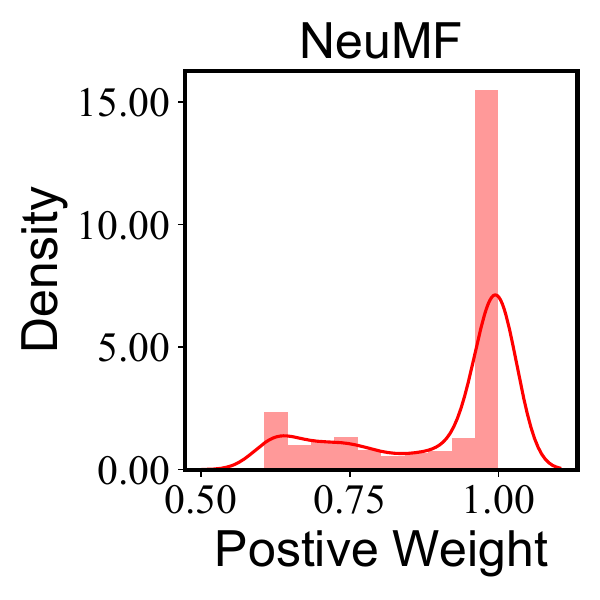}} &
 {\includegraphics[height=3.5cm]{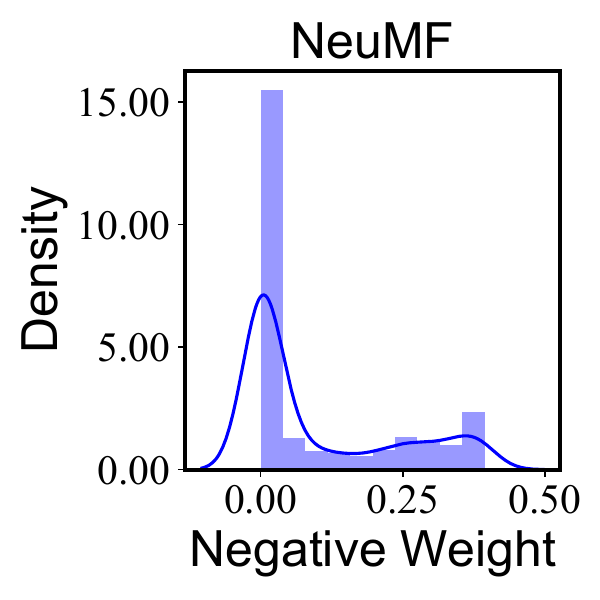}} \\
 
 {\includegraphics[height=3.5cm]{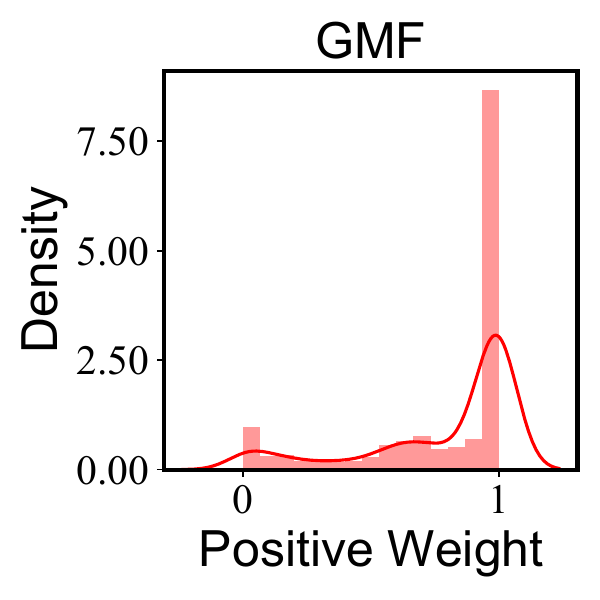}} &
 {\includegraphics[height=3.5cm]{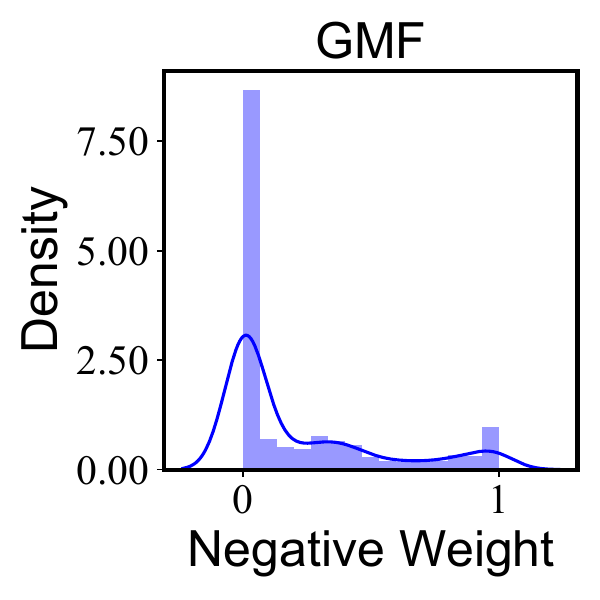}} &
 {\includegraphics[height=3.5cm]{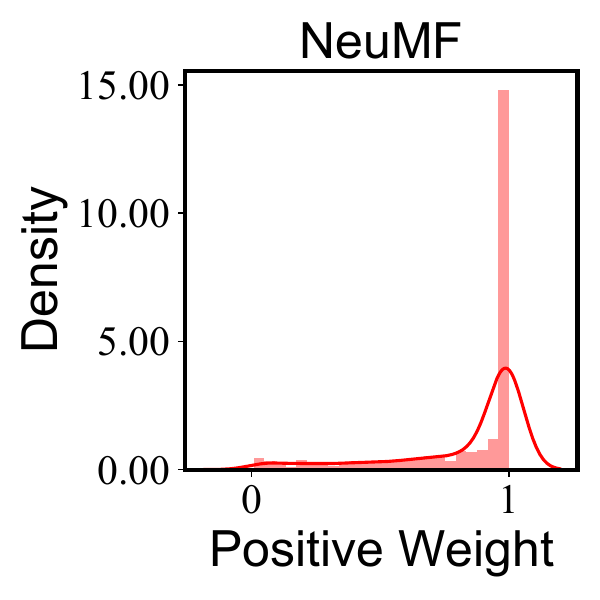}} &
 {\includegraphics[height=3.5cm]{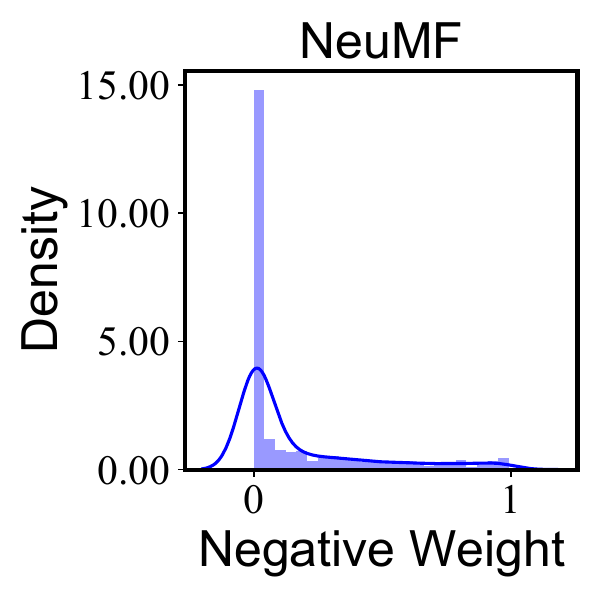}} 
\end{tabular}
	\end{center}
	\caption{Positive and negative weight distributions for dual loss on ML1M at first (up) and final (bottom) epochs. }
	\label{fig:weight_ml1m}
\end{figure*}

\begin{figure*}[!htb]
	\begin{center}
\begin{tabular}{cccc}

 {\includegraphics[height=3.5cm]{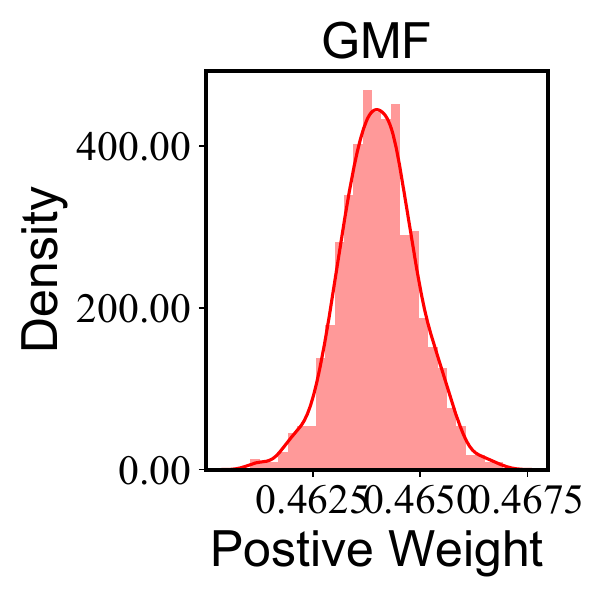}} &
 {\includegraphics[height=3.5cm]{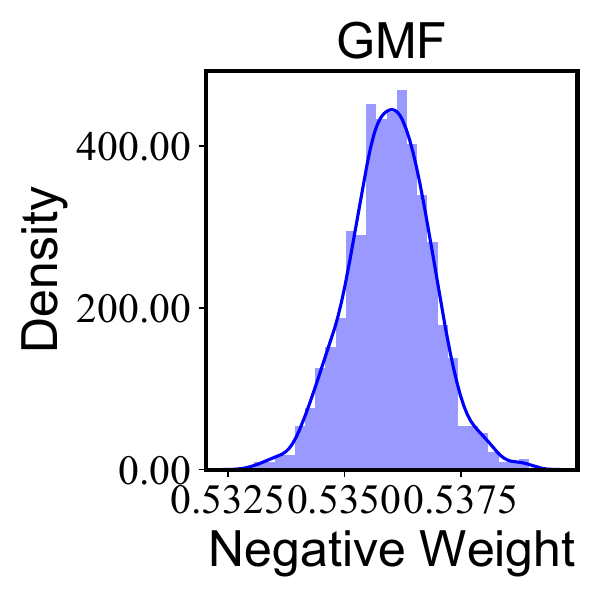}} &
 {\includegraphics[height=3.5cm]{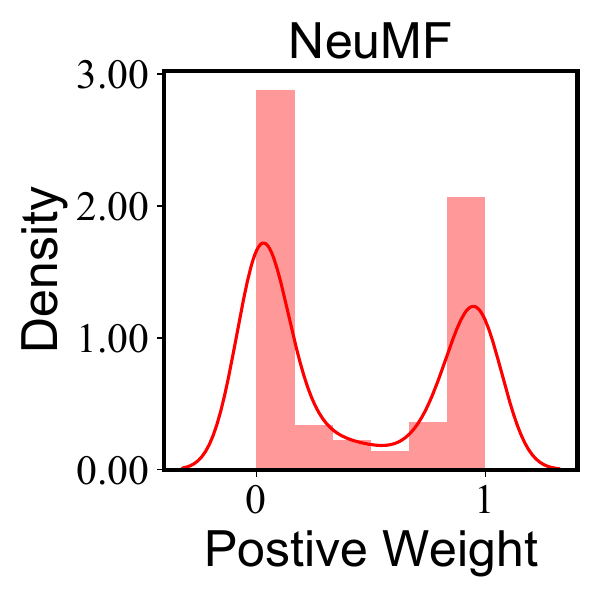}} &
 {\includegraphics[height=3.5cm]{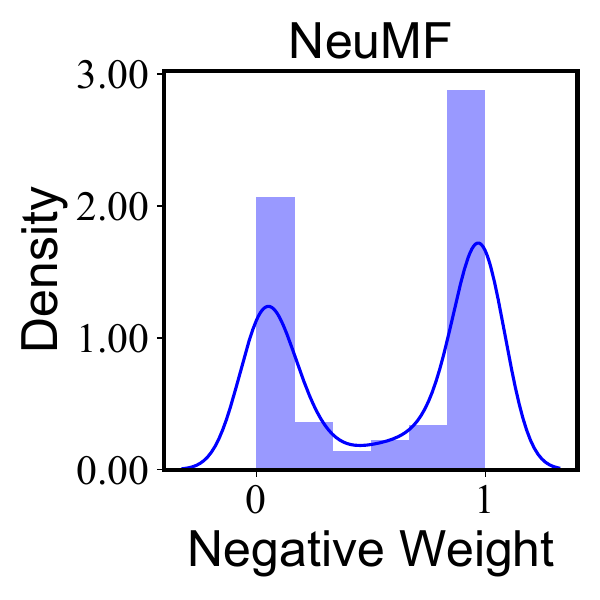}} \\
 {\includegraphics[height=3.5cm]{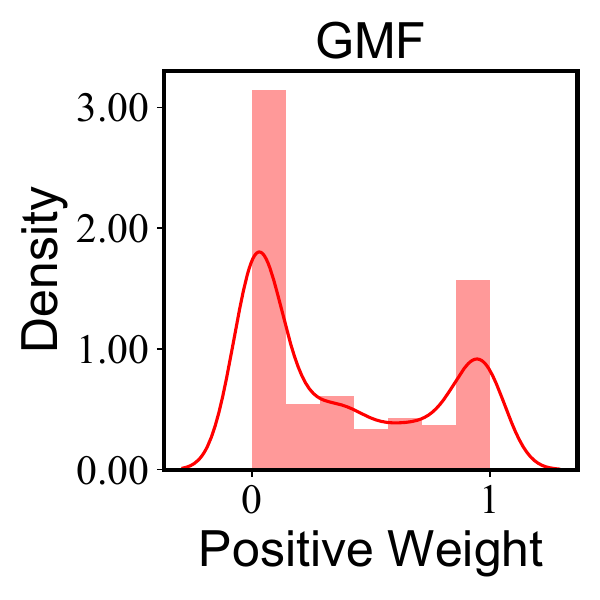}} &
 {\includegraphics[height=3.5cm]{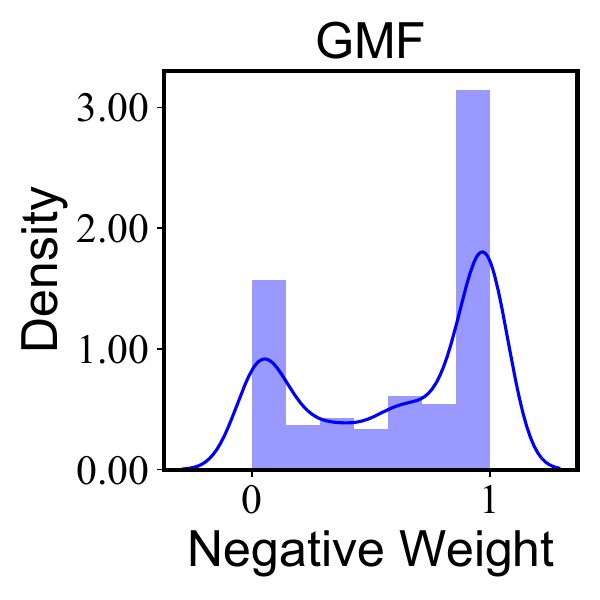}} &
 {\includegraphics[height=3.5cm]{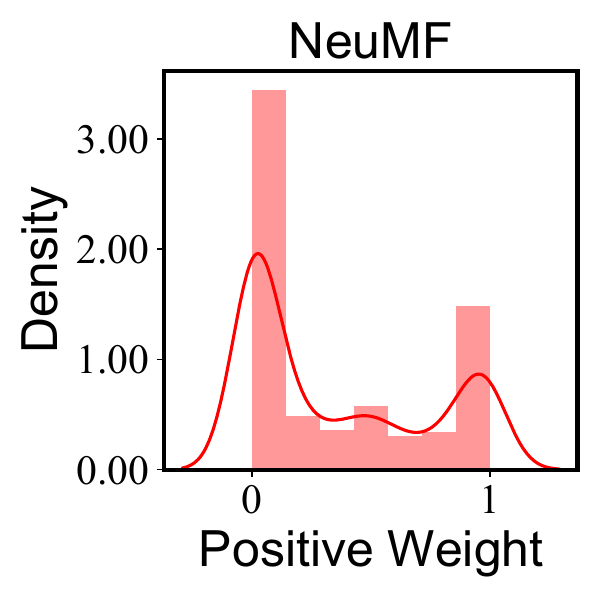}} &
 {\includegraphics[height=3.5cm]{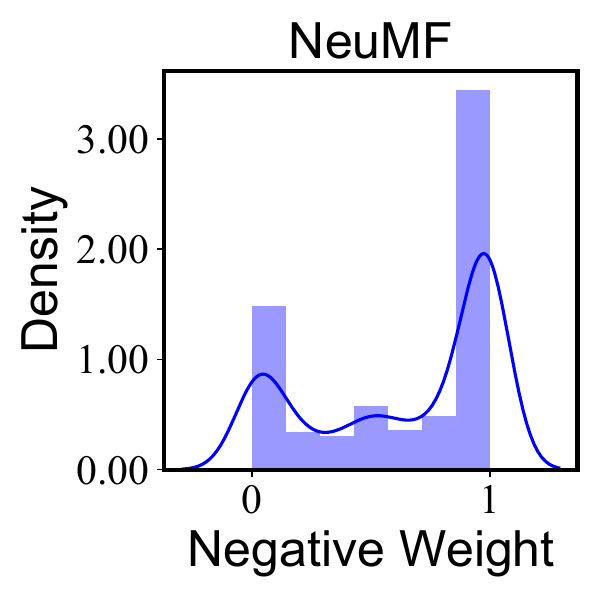}} 
\end{tabular}
	\end{center}
	\caption{Positive and negative weight distributions for dual loss on Micro Video at first (up) and final (bottom) epochs. }
	\label{fig:weight_kuaishou}
\end{figure*}

\subsubsection{\textbf{Hyper-parameter Settings}} For the two denoising strategies~\cite{Wang_2021}, we followed their default settings and verified the effectiveness of our methods under the same conditions. The embedding size and batch size of all models are set as 32 and 1,024, respectively. Besides, we adopt Adam~\cite{Adam} to optimize all the model parameters with the learning rate $\gamma$ initialized as 0.0001 and 0.00001 for labeled data on ML1M and Micro Video datasets, respectively, while the learning rate for sampled data is set as $\alpha = 0.1\gamma$. \  As for the inverse gradient, we split 90\% of the training data as training-train data, and the left is used as training-test data. The sampling rate is set as 1. The provided code has included the best hyper-parameters.

\subsection{Overall Performance (RQ1)}

The performance comparison is shown in Table~\ref{tbl:overall}, from which we have the following observations.

\begin{itemize}[leftmargin=*]
\item \textbf{Our inverse gradient performs best.} Our inverse gradient (IG) method achieves the best performance compared with four baselines and our inverse dual loss (IDL) for three metrics. 
Specifically, our IG improves the backbone sharply, which shows the ability of our proposed method to well classify the unlabeled data and achieve effective data augmentation to resolve the data sparsity problem of existing recommenders. Note that on apart from IDL and IG, GMF is better than NeuMF in general. But NeuMF with IG can significantly improve the performance, which is even better than GMF with IG. This is because NeuMF is a deep-based model, which will overfit when data is less or noise.
Besides, IG outperforms the existing negative sampling (NS) method, which means there is truly a large number of positive unlabeled data, and directly treating them all as negative feedback will confuse the model. Finally, IG also outperforms existing state-of-art denoising methods, T-CE and R-CE, showing the importance of tackling the noise from both positive and negative feedback.

\item \textbf{Inverse gradient can improve inverse dual loss to be more robust.} Inverse dual loss (IDL) only outperforms the NeuMF on ML1M dataset for AUC and GAUC, which shows the inferior robustness of IDL since it depends heavily on the training model and sampled data. Besides, it is even outperformed by GMF on ML1M and NeuMF on Micro Video, which means the poor sampling of unlabeled data will have a negative impact on the model training. These results show the significance of improving the robustness of IDL and confirm that it is necessary to exploit IG to adjust IDL.


\end{itemize}

\subsection{Annotation on Unlabeled Data (RQ2)}

To study the ability of our proposed method to annotate the unlabeled data, we visualize the distribution of weights for positive loss and negative loss in the ML1M and Micro Video datasets, respectively, as Figure~\ref{fig:weight_ml1m} and~\ref{fig:weight_kuaishou}. In specific, the upper part represents the weights at the first epoch after pre-training on the training-train data, and the bottom part represents the weights at the convergence epoch. From the figure, we can observe that: 

\begin{itemize}[leftmargin=*]

\item \textbf{Both deep and linear models are improved maximumly by our method.}
After convergence, both GMF and NeuMF can well classify the unlabeled data and capture similar patterns. 
In the ML1M dataset, most sampled instances are labeled as positive with more positive weights. In the Micro Video dataset, sampled instances are labeled in a more balanced manner. 
It means the deep model and linear model have the same upper bound on these two datasets, which is also consistent with the results of IG in overall performance where these two methods have competitive performance. This also shows the robustness of our proposed IG for improving both linear and deep models to their upper bound.
\item \textbf{Deep model annotates the unlabeled data faster than the linear model.} It is obvious that at the beginning, the linear GMF model fails to well classify the unlabeled data but the deep NeuMF model has well captured the pattern of unlabeled data, which shows the generalization ability of deep learning model~\cite{cheng2016wide}.

\end{itemize}


\subsection{Convergence Analysis (RQ3)}

\begin{figure}[t!]
	\begin{center}
\begin{tabular}{cc}

 {\includegraphics[width=.5\linewidth]{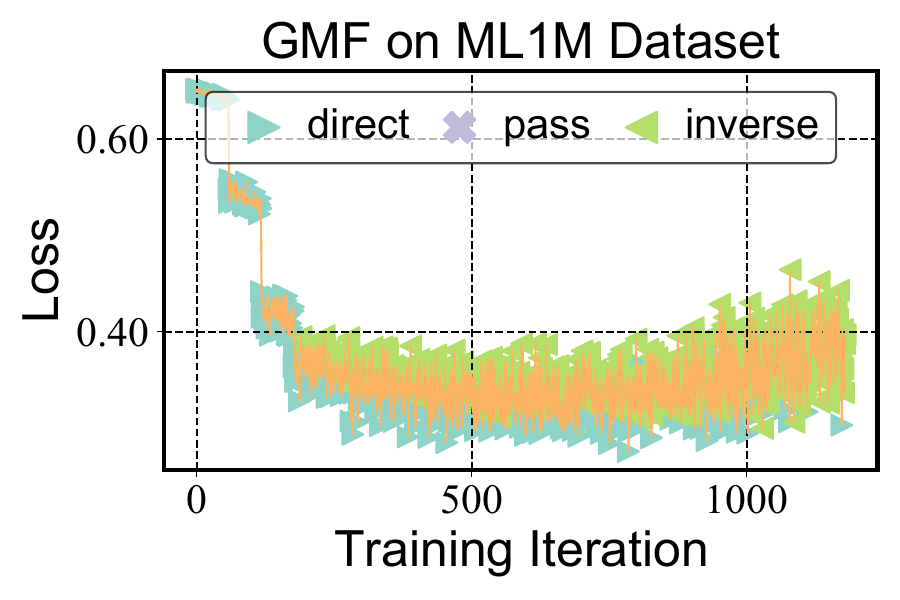}} &
 {\includegraphics[width=.5\linewidth]{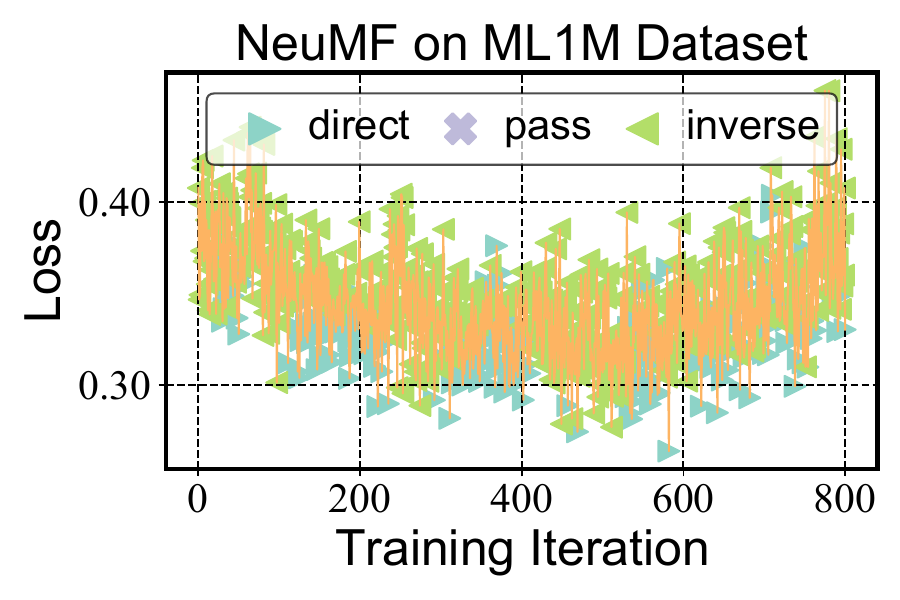}} \\
 {\includegraphics[width=.5\linewidth]{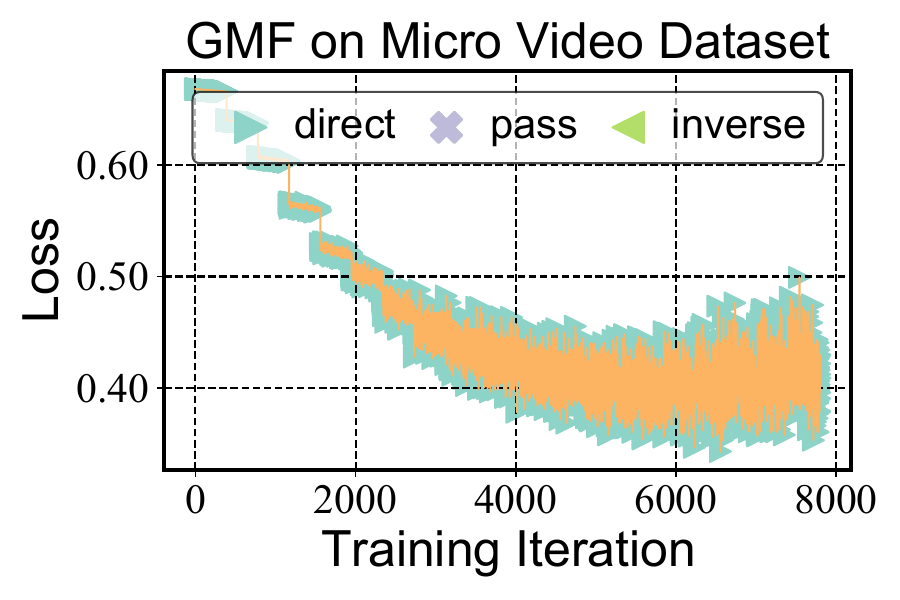}} &
 {\includegraphics[width=.5\linewidth]{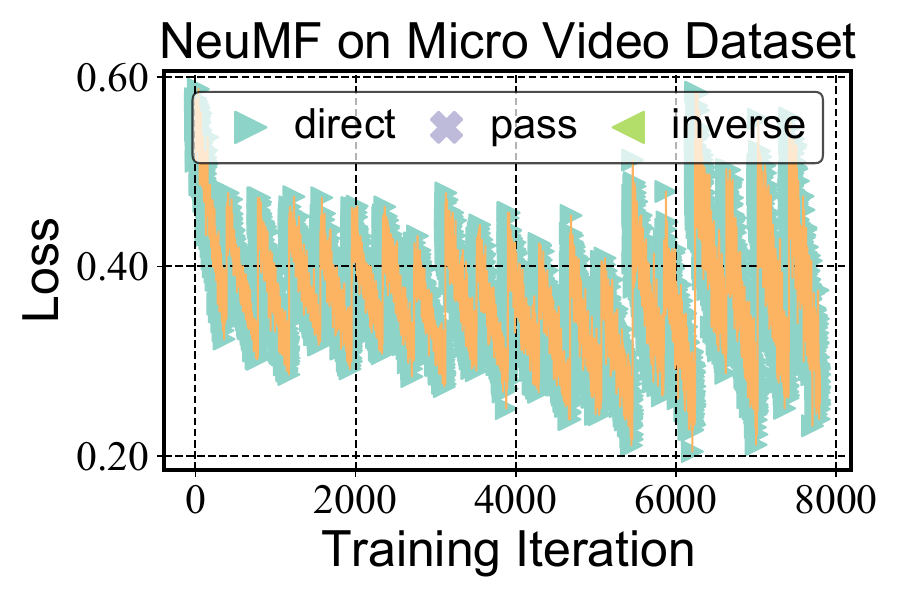}}
\end{tabular}
	\end{center}
	\caption{The loss on training-test data with the adapted gradient from the dual loss of unlabeled data. Each point of the loss curve is marked by its updated gradient direction.}
	\label{fig:loss}
\end{figure}

To investigate the convergence of our proposed inverse gradient, we also plot the loss curve of training-test data for the ML1M and Micro Video datasets on the upper and bottom parts in Figure~\ref{fig:loss}, respectively. Based on the results, we can discover that: 
\begin{itemize}[leftmargin=*]

\item \textbf{Inverse Gradient can promote convergence.}
For the ML1M dataset, at the early stage, GMF model is updated with direct gradient, then with a hybrid of direct and inverse gradients, and finally with inverse gradient. In NeuMF model, we can discover more hybrid gradients in the valley of the loss curve. This is because the test gradient is more likely to ascent at the valley where our Inverse Gradient inverses the gradient for dual loss to adjust its direction for better convergence. 

\item \textbf{Proper learning rate can prevent gradient ascent.} For the Micro Video dataset, the models are always updated with direct gradient. This is because the learning rate is relatively low here (as analyzed in the Section~\ref{sec:hyper}), leading to almost no gradient ascent problem. Most importantly, we can discover that there is no case of passing gradient in the descent procedures, which supports our analysis at Section~\ref{sec:convergence} that setting a smaller value of learning rate $\alpha$ can avoid the gradient ascent problem for inverse dual loss.

\item \textbf{Deep model converges fast but is prone to be overfitting.} 
For the ML1M dataset, the pre-training on training-train data can promote the model learning effectively at the early stage for both GMF and NeuMF models, while NeuMF will come into a fast convergence in the first epoch. However, for the Micro Video dataset, the pre-training on the training-train data will conflict with the test data, which means the deep learning-based model is prone to be overfitting~\cite{cheng2016wide} in the sparse data. 


\end{itemize}


\subsection{Hyper-parameter Study (RQ4)}\label{sec:hyper}

\begin{figure}[t!]
	\begin{center}
\begin{tabular}{cc}

 {\includegraphics[width=.49\linewidth]{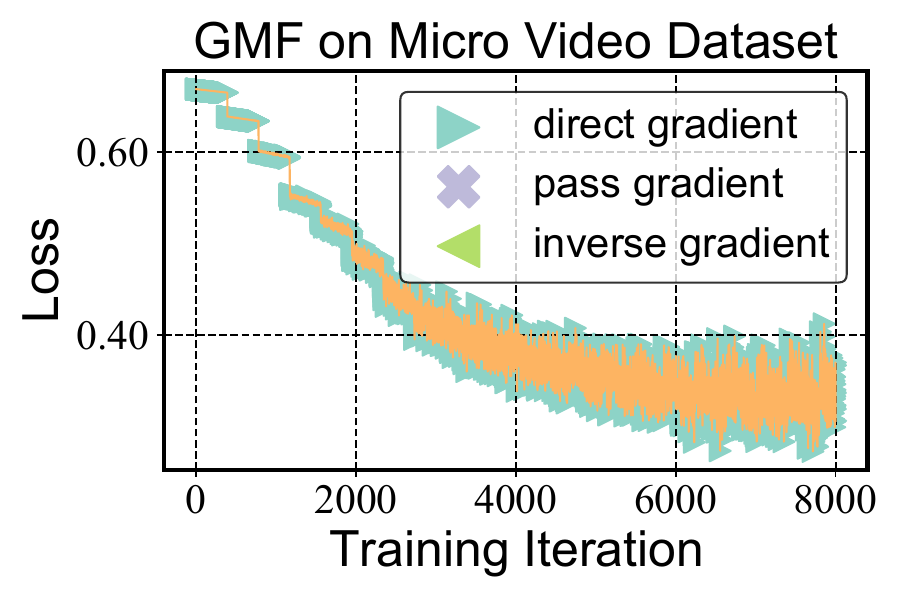}} &
 {\includegraphics[width=.49\linewidth]{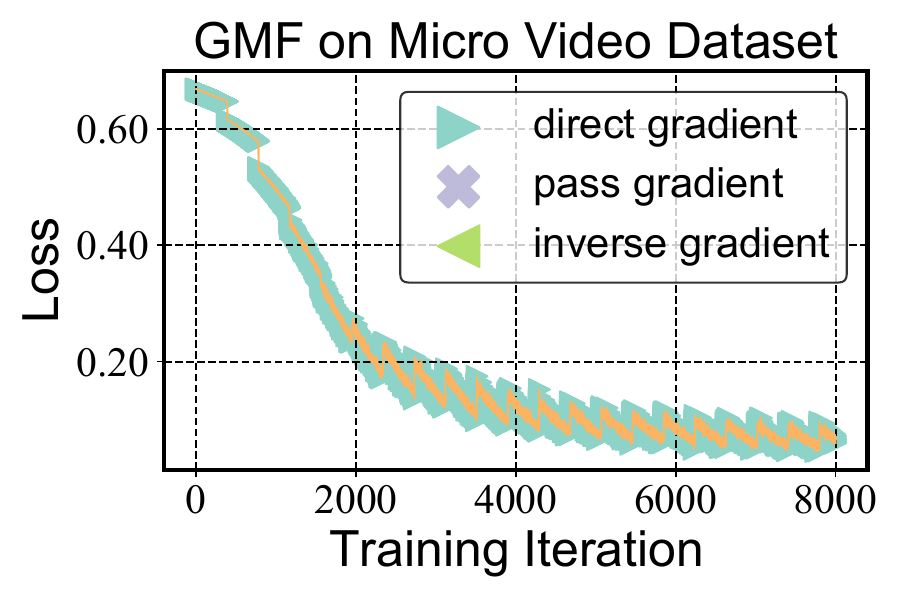}}  \\
(1) $\alpha = \gamma$ & (2) $\alpha = 10 \gamma$ \\
 {\includegraphics[width=.49\linewidth]{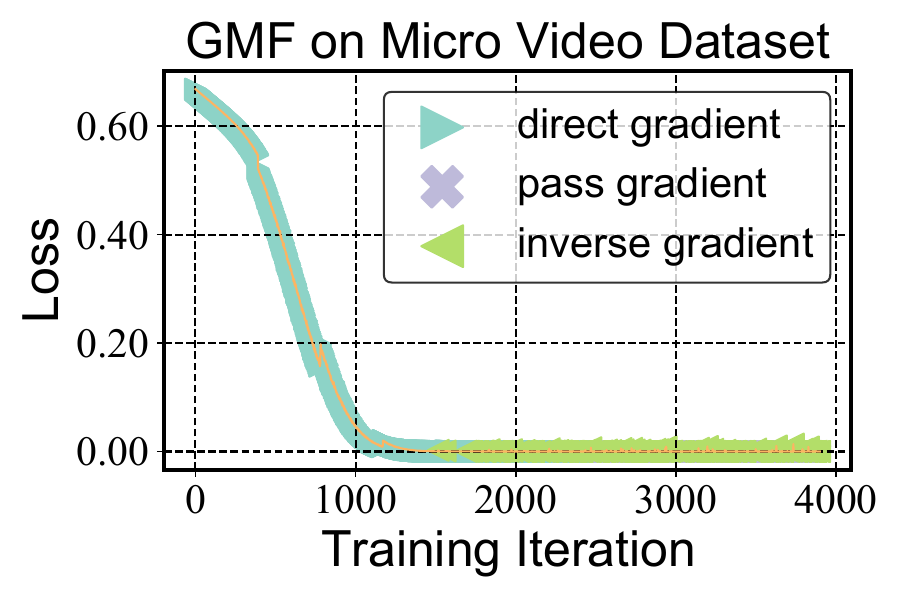}} &
 {\includegraphics[width=.49\linewidth]{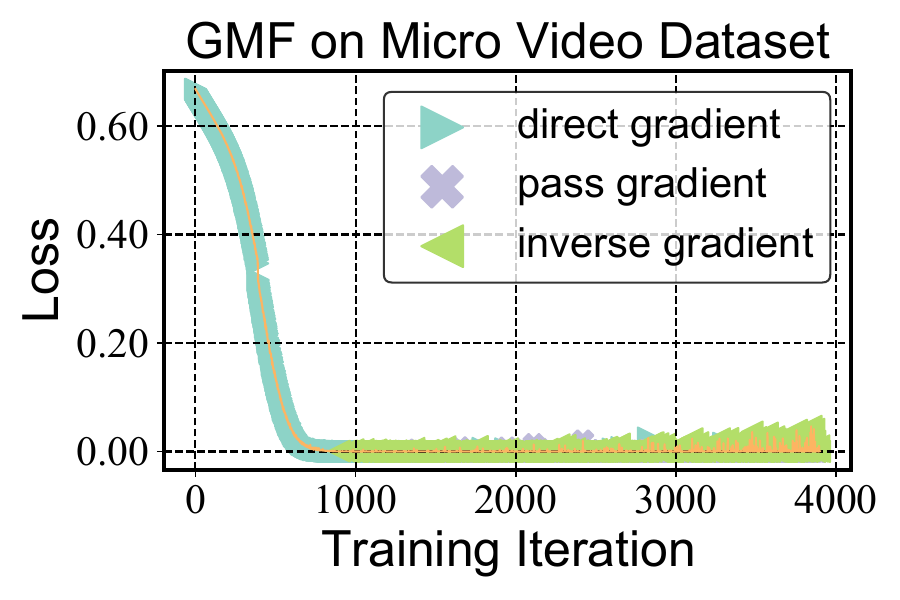}}  \\
 (3) $\alpha = 50 \gamma$ & (4) $\alpha = 100 \gamma $
\end{tabular}
	\end{center}
 \vspace{-0.2cm}
	\caption{The loss of GMF model on training-test data with different learning rates for adapted gradient under Micro Video dataset. Each point of the loss curve is marked by its updated gradient direction. }
	\label{fig:loss_lr}
 \vspace{-0.2cm}
\end{figure}

As discussed in Section~\ref{sec:convergence}, we can avoid the gradient ascent problem by setting the learning rate for inverse dual loss smaller than that for test loss, i.e., $\alpha < \gamma$ in Algorithm~\ref{alg:ige}.
To experimentally study this convergence analysis at Section~\ref{sec:convergence} and investigate the impact of the learning rate for convergence, we vary the learning rate for inverse dual loss with 1, 10, 50, and 100 times the learning rate for test loss.
 Here we study the loss curve of GMF on Micro Video as Figure~\ref{fig:loss_lr}, where we can discover that:

\begin{itemize}[leftmargin=*]
    
\item \textbf{Smaller learning rate for inverse dual loss can avoid gradient ascent.}
When $\alpha$ is smaller than 10 times $\gamma$, the gradients are often direct gradients. However, when $\alpha$ grows up to 50 times $\gamma$, the inverse gradients appear. Moreover, when $\alpha$ grows up to 100 times $\gamma$, there even appear pass gradients which means the occurrence of gradient ascent. Thus it is consistent to our analysis at Section~\ref{sec:convergence} that we can avoid the gradient ascent problem by limiting the learning rate for inverse dual loss according to the learning rate for test loss.

\item \textbf{Greater learning rate for inverse dual loss can speed up the convergence but more fluctuation.} 
With the growth of learning rate $\alpha$, the loss convergence becomes faster. However, it also results in fluctuation as there appear more inverse gradients and pass gradients. This observation is consistent to our analysis of Section~\ref{sec:convergence} for convergence.

\end{itemize}

\section{Related Work}
\vspace{0.2cm}
\textbf{Implicit Feedback with Negative Sampling.} Existing recommenders are generally based on implicit feedback data, where the collected data is often treated as positive feedback, and negative sampling~\citep{rendle2012bpr, he2017neural, chen2019social} is exploited to balance the lack of negative instances. However, the negative sampling strategy will introduce noise because there are some positive unlabeled~\cite {elkan2008learning, bekker2019beyond, saito2020unbiased} data in the sampled instances. To improve existing implicit feedback recommendation, the identification of negative experiences~\citep{fox2005evaluating, kim2014modeling} has grabbed the researchers' attention.
However, these methods collect either the various user feedback (e.g., dwell time~\citep{kim2014modeling} and skip~\citep{wen2019leveraging}) or the item characteristics~\citep{lu2018between}, requiring additional feedback and manual labeling, e.g., users are supposed to actively provide their satisfaction.
Besides, the evaluation of items relies heavily on manual labeling and professional knowledge~\cite {lu2018between}. Thus in practice, these methods are too expensive to implement in real-world recommenders. 
In addition, hard negative sampling is adopted to improve the negative sampling~\citep{DNS, RNS, NEURIPS2020_0c7119e3}. However, with fewer false positive samples, the hard negative instances also bring more false-negative samples. Our meta-learning method elegantly annotates the unlabeled instances based on the sparsely labeled instances.

\vspace{0.2cm}

\textbf{Denoising Recommender Systems.} One intuitive approach to reduce noise is to directly include more accurate feedback~\citep{liu2010understanding, yang2012exploiting}, such as dwell time~\citep{yi2014beyond} and skip~\citep{wen2019leveraging}). However, forcefully requiring additional feedback from users may harm user experiences. 
To address this problem, DenoisingRec~\cite {Wang_2021} achieves denoising recommendation for implicit feedback without any additional data. More specifically, they perform denoising on the false positive instances via truncating or reweighting the samples with larger loss. However, they only consider the positive feedback without further addressing the noise brought by negative sampling. Our work considers the sampled instances as possibly positive and negative, then achieve denoising data augmentation from both positive and negative perspectives.

\section{Conclusions and Future Work}

In this paper, we proposed a novel method that automatically annotated the unlabeled data and adjusted the false annotated labels.
 Such exploration not only addressed the unavoidable noise brought by widely used negative sampling but also improved the current denoising recommenders.
Specifically, we proposed inverse learning from both loss and gradient perspectives. The first one was the Inverse Dual Loss that assumed the sampled data to be possibly positive or negative and automatically annotated them. 
If the positive loss was greater than the negative loss (difficult to label the data as positive), the Inverse Dual Loss would inversely assign more weights to the negative loss and vice versa. 
Since the Inverse Dual Loss depended heavily on the current training model and the quality of sampled data, we further proposed Inverse Gradient which made Inverse Dual Loss more robust by adjusting the gradient for those falsely annotated instances. 
We designed a meta-learning method with the training data split into training-train data and training-test data. 
The model was first pre-trained on the training-train data. 
Then the pre-trained model would explore updating with the gradient or the additive inverse of the gradient, or even did not update, determined by the training-test data.

As for future work, we plan to apply our inverse learning with more recommendation models as the backbones to further verify the generalization of our proposed methods.

\clearpage
\bibliographystyle{ACM-Reference-Format}
\bibliography{sample-base}
\clearpage
\appendix 
\section{APPENDIX}

\subsection{Proof of Theorem 1}\label{appendix::proof}
\begin{theorem1} \label{def:inverse_grad}
	Given learning rate $\alpha \in \mathbb{R}, \alpha \neq 0$, assume the temporal model parameters updated by the direct gradient and inverse gradient, respectively, are as $\boldsymbol{\theta}^d = \boldsymbol{\theta} - \alpha \circ \nabla \mathcal{L}^{dual}_{\mathcal{D}^u}(\boldsymbol{\theta})$ and $\boldsymbol{\theta}^i = \boldsymbol{\theta} + \alpha \circ \nabla \mathcal{L}^{dual}_{\mathcal{D}^u}(\boldsymbol{\theta})$. Then, the relationship between the loss of them and the model with parameter $\boldsymbol{\theta}$ on data instance $\left(u, i, y_{u i}^{*}\right) \in \mathcal{D}^l$ will be either $\mathcal{L}_{\mathcal{D}^l}(\boldsymbol{\theta}^d) > \mathcal{L}_{\mathcal{D}^l}(\boldsymbol{\theta}) >  \mathcal{L}_{\mathcal{D}^l}(\boldsymbol{\theta}^i) $ or $ \mathcal{L}_{\mathcal{D}^l}(\boldsymbol{\theta}^d) < \mathcal{L}_{\mathcal{D}^l}(\boldsymbol{\theta}) < \mathcal{L}_{\mathcal{D}^l}(\boldsymbol{\theta}^i)$.
\end{theorem1}

\begin{proof} 
To simplify the problem, from a stochastic perspective with one instance $\left(u, i, y_{u i}^{*}\right)$. Our target is to satisfy:
\begin{equation}\label{eq:target1}
	\mathcal{L}(\boldsymbol{\theta}^d) > \mathcal{L}(\boldsymbol{\theta}) >  \mathcal{L}(\boldsymbol{\theta}^i),
\end{equation}
or
\begin{equation}\label{eq:target2}
\mathcal{L}(\boldsymbol{\theta}^d) < \mathcal{L}(\boldsymbol{\theta}) < \mathcal{L}(\boldsymbol{\theta}^i),
\end{equation}
where $\mathcal{L}(\boldsymbol{\theta}^d), \mathcal{L}(\boldsymbol{\theta}), \mathcal{L}(\boldsymbol{\theta}^i)$ are the loss functions for the models with parameters $\boldsymbol{\theta}^d, \boldsymbol{\theta}, \boldsymbol{\theta}^i$. To simplify the proof procedure, we define the prediction function as the hypothesis function of the well known logistic regression~\citep{hosmer2013applied}\footnote{\url{https://www.coursera.org/learn/machine-learning}}:
\begin{equation}\label{eq:hyposis}
	\hat{y}^{\boldsymbol{\theta}}_{u i}=f(u, i | {\boldsymbol{\theta}}) = f(\boldsymbol{x}_{u,i} | {\boldsymbol{\theta}}) = \frac{1}{1+e^{-\boldsymbol{\theta} \boldsymbol{x}_{u,i}}},
\end{equation}
where $\boldsymbol{x}_{u,i}$ is the input feature under the interaction between user $u$ and item $i$.
 Given learning rate $\alpha \in \mathbb{R}, \alpha \neq 0$ with $\boldsymbol{\theta}^d = \boldsymbol{\theta} - \alpha \circ \nabla \mathcal{L}^{dual}_{\mathcal{D}^u}(\boldsymbol{\theta})$ and $\boldsymbol{\theta}^i = \boldsymbol{\theta} + \alpha \circ \nabla \mathcal{L}^{dual}_{\mathcal{D}^u}(\boldsymbol{\theta})$ as the temporal model parameters updated by the direct gradient and inverse gradient, respectively, we can have:
\begin{equation}
\boldsymbol{\theta}^d \boldsymbol{x}_{u,i} =\boldsymbol{\theta} \boldsymbol{x}_{u,i}  -  \alpha \circ \nabla \mathcal{L}^{dual}_{\mathcal{D}^u}(\boldsymbol{\theta})\boldsymbol{x}_{u,i} 
\end{equation}
\begin{equation}
	\boldsymbol{\theta}^i \boldsymbol{x}_{u,i} =\boldsymbol{\theta} \boldsymbol{x}_{u,i}  +  \alpha \circ \nabla \mathcal{L}^{dual}_{\mathcal{D}^u}(\boldsymbol{\theta})\boldsymbol{x}_{u,i} 
\end{equation}

Assume $\alpha \circ \nabla \mathcal{L}^{dual}_{\mathcal{D}^u}(\boldsymbol{\theta})\boldsymbol{x}_{u,i} < 0$ then we have:
\begin{equation}\label{eq:assumption}
	\boldsymbol{\theta}^d \boldsymbol{x}_{u,i} > \boldsymbol{\theta} \boldsymbol{x}_{u,i} > 	\boldsymbol{\theta}^i \boldsymbol{x}_{u,i}
\end{equation}
Review the loss function from a stochastic perspective with one instance $\left(u, i, y_{u i}^{*}\right)$:
\begin{equation}\label{eq:loss_stochastic}
	\mathcal{L}(\boldsymbol{\theta}) = -  y_{u i}^{*} \log \left(\hat{y}^{\boldsymbol{\theta}}_{u i}\right) - \left(1-y_{u i}^{*}\right) \log \left(1-\hat{y}^{\boldsymbol{\theta}}_{u i}\right) .
\end{equation}
Suppose $\left(u, i, y_{u i}^{*}\right)$ is positive instance with $y_{u i}^{*} = 1$, then we can have:
\begin{equation}\label{eq:loss_stochastic_pos}
	\mathcal{L}(\boldsymbol{\theta}) = -  \log \left(\hat{y}^{\boldsymbol{\theta}}_{u i}\right),
\end{equation} 
which is a decreasing function towards predicted probability $\hat{y}^{\boldsymbol{\theta}}_{u i}$. 

$\mathcal{L}(\boldsymbol{\theta})$ will decrease towards $	\hat{y}^{\boldsymbol{\theta}}_{u i}$ and further decrease towards $\boldsymbol{\theta} \boldsymbol{x}_{u,i}$. Base on the assumption of Eqn.\eqref{eq:assumption}, we can have:
\begin{equation}\label{eq:result}
\mathcal{L}(\boldsymbol{\theta}^d) < \mathcal{L}(\boldsymbol{\theta}) < \mathcal{L}(\boldsymbol{\theta}^i),
\end{equation}
which satisfies the target of Eqn.\eqref{eq:target2}. 
Besides, similarly, if $y^{*}_{ui} = 0$, we can have: $	\mathcal{L}(\boldsymbol{\theta}^d) > \mathcal{L}(\boldsymbol{\theta}) > \mathcal{L}(\boldsymbol{\theta}^i)$. If $\alpha \circ \nabla \mathcal{L}^{dual}_{\mathcal{D}^u}(\boldsymbol{\theta})\boldsymbol{x}_{u,i} > 0$, we can have similar conclusion.
That is to say, whatever cases, targets of Eqn.\eqref{eq:target1} and Eqn.\eqref{eq:target2} will be satisfied.

The case $\alpha \circ \nabla \mathcal{L}^{dual}_{\mathcal{D}^u}(\boldsymbol{\theta})\boldsymbol{x}_{u,i} = 0$ will result in gradient vanishing and modern machine learning approach often randomly initialize the  features to avoid such case.
\end{proof}


\subsection{Baselines}\label{appendix::baselines}
The details of training strategies are as below.
\begin{itemize}[leftmargin=*]
\item Traditional strategies: standard loss without sampling and with negative sampling (NS)~\cite{rendle2012bpr, he2017neural, chen2019social}.
\item Denoising strategies: T-CE~\cite{Wang_2021} of DenoisingRec that truncates the loss for false-positive instances and R-CE~\cite{Wang_2021} of DenoisingRec weighs less on the loss for false-positive instances; Our proposed inverse dual loss that weighs more on the label with a smaller loss (true-positive or true-negative instances).
\end{itemize}
Beside, we also illustrate the backbones here.
\begin{itemize}[leftmargin=*]
     \item GMF~\cite{he2017neural}: A variant of matrix factorization with the element-wise product and a linear neural layer as the interaction function instead of the inner product. 
    \item NeuMF~\cite{he2017neural}: A combination of GMF and Multi-Layer Perceptron.
\end{itemize}

\subsection{Hard Negative Sampling Comparison (RQ5)}\label{appendix::hardNegCom}
\begin{table}[t!]

\caption{Performance comparisons towards hard negative sampling with GMF and NeuMF backbones on two datasets. Bold and underline refer to the best and second best result, respectively.}\label{tbl:hard_nega}
\setlength\tabcolsep{1pt}

\centering
\begin{tabular}{cccccccc}
\hline
                                 &                                   & \multicolumn{3}{c}{\textbf{ML1M}}                                                             & \multicolumn{3}{c}{\textbf{Micro Video}}            \\ \cline{3-8} 
\multirow{-2}{*}{\textbf{Model}} & \multirow{-2}{*}{\textbf{Method}} & \textbf{NDCG}                 & \textbf{AUC}                  & \textbf{GAUC}                 & \textbf{NDCG}   & \textbf{AUC}    & \textbf{GAUC}   \\ \hline
                                 & \textbf{None}                     & 0.9285                        & {\ul 0.7671}                  & 0.6919                        & {\ul 0.7365}    & {\ul 0.8024}    & {\ul 0.7558}    \\ \cline{2-8} 
                                 & \textbf{NS}                       & {\ul 0.9400}                  & 0.7639                        & {\ul 0.7200}                  & 0.7049          & 0.7802          & 0.7247          \\ \cline{2-8} 
                                 & \textbf{DNS}                      & {\color[HTML]{333333} 0.9178} & {\color[HTML]{333333} 0.6290} & {\color[HTML]{333333} 0.6290} & 0.6878          & 0.6310          & 0.6356          \\ \cline{2-8} 
                                 & \textbf{SRNS}                     & {\color[HTML]{333333} 0.9176} & {\color[HTML]{333333} 0.6253} & {\color[HTML]{333333} 0.6293} & 0.6799          & 0.6293          & 0.6302          \\ \cline{2-8} 
                                 & \textbf{IDL}                      & 0.8996                        & 0.7304                        & 0.6400                        & 0.7272          & 0.7858          & 0.7335          \\ \cline{2-8} 
\multirow{-6}{*}{\textbf{GMF}}   & \textbf{IG}                       & \textbf{0.9521}               & \textbf{0.8318}               & \textbf{0.7642}               & \textbf{0.7773} & \textbf{0.8033} & \textbf{0.7593} \\ \hline
                                 & \textbf{None}                     & 0.9214                        & 0.7524                        & 0.6856                        & {\ul 0.6649}    & {\ul 0.7504}    & {\ul 0.6916}    \\ \cline{2-8} 
                                 & \textbf{NS}                       & {\ul 0.9298}                  & 0.7495                        & {\ul 0.7088}                  & 0.6350          & 0.7191          & 0.6689          \\ \cline{2-8} 
                                 & \textbf{DNS}                      & 0.9168                        & 0.6277                        & 0.6283                        & 0.6203          & 0.6274          & 0.6261          \\ \cline{2-8} 
                                 & \textbf{SRNS}                     & 0.9172                        & 0.6282                        & 0.6284                        & 0.6212          & 0.6296          & 0.6295          \\ \cline{2-8} 
                                 & \textbf{IDL}                      & 0.9212                        & {\ul 0.7962}                  & 0.6984                        & 0.6466          & 0.7174          & 0.6647          \\ \cline{2-8} 
\multirow{-6}{*}{\textbf{NeuMF}} & \textbf{IG}                       & \textbf{0.9449}               & \textbf{0.8253}               & \textbf{0.7569}               & \textbf{0.7809} & \textbf{0.8198} & \textbf{0.7689} \\ \hline
\end{tabular}
\end{table}
To study the performance comparison of our proposed method with SOTA hard negative sampling methods, we compare our method with two typical baselines as below.
\begin{itemize}[leftmargin=*]
     \item DNS~\citep{DNS}: Dynamic Negative Sampling (DNS) samples unlabeled instances as negatives and picks the hard instances with high predicted scores.
    \item SRNS~\citep{NEURIPS2020_0c7119e3}: SRNS further selects true negatives with high variance to improve DNS.
\end{itemize}

The results of performance comparison with hard negative sampling methods are as Table~\ref{tbl:hard_nega}, where we can discover that the hard negative sampling methods truly perform worse in our setting where both ground-truth positive and negative samples are tested. Here hard negative sampling methods are even outperformed by the negative sampling and pure method without sampling, which means they truly increase the false-negative instances as discussed in Section~\ref{sec:intro}.

\subsection{Datasets and Pre-processing}\label{appendix::dataset}
We introduce the details of these datasets, including the pre-processing steps, as follows.

\begin{table}[t]
\caption{Data statistics for processed Micro Video dataset and ML1M dataset.}\label{tbl:data}
\centering
\begin{tabular}{cccc}
\hline
\multicolumn{2}{c}{\textbf{Dataset}} & \textbf{Micro Video} & \textbf{ML1M} \\ \hline
\multicolumn{2}{c}{\textbf{Users}} & 37,692 & 6,041 \\ \hline
\multicolumn{2}{c}{\textbf{Items}} & 131,690 & 3,953 \\ \hline
\multirow{3}{*}{\textbf{Feedback}} & \textbf{Positive} & 4,915,745 & 836,478 \\ \cline{2-4} 
 & \textbf{Negative} & 4,546,747 & 163,731 \\ \cline{2-4} 
 & \textbf{Total} & 9,462,492 & 1,000,209 \\ \hline
\multicolumn{2}{c}{\textbf{Density}} & 0.19\% & 4.19\% \\ \hline
\end{tabular}
\end{table}
The public ML1M dataset is published at\footnote{\url{https://grouplens.org/datasets/movielens/1m/}}, and we also have uploaded the processed dataset on the link of the code and the supplementary material. The statistics of our adopted Micro Video dataset and ML1M dataset are as Table~\ref{tbl:data}. We will put the Micro Video dataset public to benefit the community.
\begin{table}[!htb]
\caption{Interaction statistics for Micro Video dataset.}\label{tbl:data_before}
\setlength\tabcolsep{2pt}
\centering
\begin{tabular}{cccccc}
\hline
\textbf{Dataset}     & \textbf{Like} & \textbf{Hate} & \textbf{Total} & \textbf{Like/Total} & \textbf{Hate/Total} \\ \hline
\textbf{Micro Video} & 382,570       & 6,157         & 8,008,965      & 4.78\%              & 0.08\%              \\ \hline
\end{tabular}
\end{table}

\begin{figure}[t]
	\begin{center}
\begin{tabular}{cc}

 {\includegraphics[width=.5\linewidth]{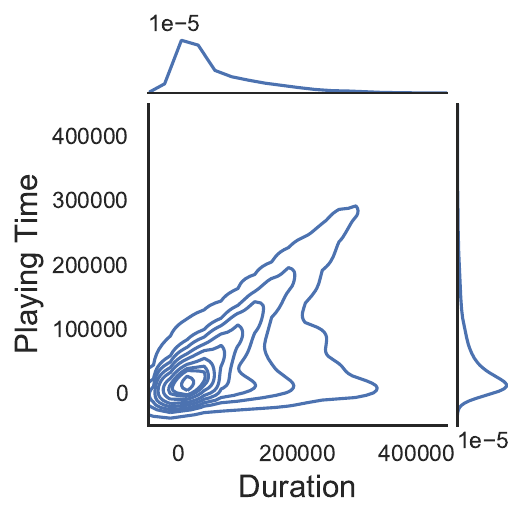}} &
 {\includegraphics[width=.5\linewidth]{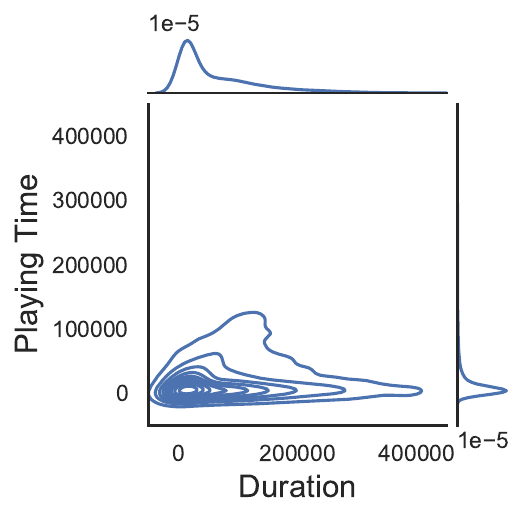}} 
\end{tabular}
	\end{center}
	\caption{Joint distribution of playing time and duration for like (left) and hate (right). }
	\label{fig:finsh_rate}
\end{figure}

\para{Micro Video} This dataset is collected from one of the largest Micro Video platforms in China, where user behaviors such as playing time, like, and hate are recorded. The data is downsampled from September 11 to September 22, 2021. Users are passive to receive the recommended videos here, and there is extremely limited active feedback, such as like and hate as shown in Table~\ref{tbl:data_before}. That is to say, we have extremely limited reliable feedback in this data, which is very challenging in modern industries. We take the active feedback of like and hate to analyze the regular pattern of playing time and duration of each video which are included in each interaction. From Figure~\ref{fig:finsh_rate}, we can discover that the users' like and hate behaviors are related to the finish rate of playing time (user's playing time towards a certain item) to be divided by duration (item's total time): when the users like the video, they are more likely to finish watching it, and vice versa. Thus we treat the finish rate greater than 80\% and less than 20\% as positive and negative feedback, respectively. Another way to classify the positive and negative feedback is according to the playing time, and we also report the results by treating the playing time over the upper quartile and under the lower quartile as positive and negative feedback, respectively. From Table~\ref{tbl:feedback_cla}, we can observe that the finish rate is more suitable for pattern capturing, and we adopt such a way as feedback processing in experimental evaluation.

\begin{table}[t]
	\caption{Performance comparison of GMF on Micro Video dataset based on feedback classification by finish rate and playing time. }
	\label{tbl:feedback_cla}
\centering
\begin{tabular}{ccc}
\hline
\textbf{Metric}       & \textbf{AUC}    & \textbf{GAUC}   \\ \hline
Finish Rate  & 0.8024 & 0.7558 \\ \hline
Playing Time & 0.751  & 0.528  \\ \hline
\end{tabular}
\end{table}

\para{ML1M}\footnote{\url{https://grouplens.org/datasets/movielens/1m/}}  This is a widely used public movie dataset in the recommendation. The rating score in ML1M ranges from 1 to 5, and we treat the rating score over 3 and under 2 as positive and negative feedback, respectively, following DenoisingRec~\citep{Wang_2021}.

Besides, we split 60\%, 20\%, and 20\% of the data as training, validation, and test data for these two datasets.

\subsection{Implementation Details}
All the models are implemented based on Python with a Pytorch\footnote{\url{https://pytorch.org/}} framework based on the repository DenoisingRec\footnote{\url{https://github.com/WenjieWWJ/DenoisingRec}}. The environment is as below.
\begin{itemize}
    \item Anaconda 3
\item python 3.7.3
\item pytorch 1.4.0
\item numpy 1.16.4
\end{itemize}

\end{document}